\documentclass[lettersize,journal]{IEEEtran}
\usepackage{amsmath,amsfonts}
\usepackage{algorithmic}
\usepackage{array}
\usepackage[caption=false,font=footnotesize,labelfont=sf,textfont=sf]{}
\usepackage{textcomp}
\usepackage{stfloats}
\usepackage{url}
\usepackage{verbatim}
\usepackage{graphicx}
\usepackage{subcaption}
\def\BibTeX{{\rm B\kern-.05em{\sc i\kern-.025em b}\kern-.08em
    T\kern-.1667em\lower.7ex\hbox{E}\kern-.125emX}}
\usepackage{balance}

\usepackage{multirow}
\usepackage{diagbox}
\usepackage{optidef}
\usepackage{bm}
\usepackage{hyperref}
\usepackage{cite}
\usepackage{xcolor}

\usepackage{tabularx}
\usepackage{booktabs}
\usepackage{hhline}

\usepackage{rotating}

\usepackage[labelsep=period]{caption}

\usepackage{amssymb,stmaryrd}

\usepackage{kotex}

\renewcommand{\arraystretch}{1.2}  

\allowdisplaybreaks

\begin{document}

\title{
evS$^{2}$CP: Real-time Simultaneous Speed and Charging Planner for Connected Electric Vehicles
}

\author{Minwoo Gwon, Jiwon Kim, Seungjun Yoo and Kwang-Ki K. Kim${}^{*}$
\thanks{The authors are with the Department of Electrical and Computer Engineering at Inha University, Incheon 22212, Korea.}
\thanks{${}^{*}$Corresponding author (Email: {\tt kwangki.kim@inha.ac.kr}).}
\thanks{This research was partly supported by the Basic Science Research Program through the National Research Foundation of Korea (NRF), funded by the Ministry of Education under Grant NRF-2022R1F1A1076260. It was also supported in part by the Korea Institute for Advancement of Technology (KIAT) grant funded by the Korea Government (MOTIE) (P0017124, The Competency Development Program for Industry Specialist), and in part by the BK21 Four Program, funded by the Ministry of Education (MOE, Korea) and the National Research Foundation of Korea (NRF).}
}

\maketitle


\begin{abstract}
This paper presents \textbf{evS$^{2}$CP}, an optimization-based framework for simultaneous speed and charging planning designed for connected electric vehicles (EVs). With EVs emerging as competitive alternatives to internal combustion engine vehicles, overcoming challenges such as limited charging infrastructure is crucial. \textbf{evS$^{2}$CP} addresses these issues by minimizing the travel time, charging time, and energy consumption, providing practical solutions for both human-operated and autonomous vehicles. This framework leverages V2X communication to integrate essential EV planning data, including route geometry, real-time traffic conditions, and charging station availability, while simulating dynamic driving environments using open-web API services. The speed and charging planning problem was initially formulated as a nonlinear programming model, which was then convexified into a quadratic programming model without charging-stop constraints. Additionally, a mixed-integer programming approach was employed to optimize charging station selection and minimize the frequency of charging events. A mixed-integer quadratic programming implementation exhibited exceptional computational efficiency and scalability, effectively solving trip plans over distances exceeding 700 km in a few seconds. Simulations conducted using open-source and commercial solvers validated the framework’s near-global optimality, demonstrating its robustness and feasibility for real-world applications in connected EV ecosystems.
\end{abstract}




\section{Introduction}
\label{sec:intro}
With stricter global environmental regulations, the automobile industry faces increasingly stringent standards for greenhouse-gas emissions, particularly for CO\(_2\) and NO$_{\textrm{\small x}}$. Thus, automotive manufacturers are accelerating the development of electric vehicles (EVs), such as battery EVs, hybrid EVs, and fuel-cell EVs, as sustainable alternatives to internal combustion engine vehicles. However, the underdeveloped charging infrastructure remains a major obstacle to the widespread adoption of these technologies.

To address the challenge of \emph{range anxiety}, recent studies have explored eco-driving approaches that leverage optimal control methodologies to enhance energy efficiency and extend driving range. For example, methods have been proposed to minimize energy consumption by selecting efficient routes and maintaining optimal speed profiles~\cite{watzenig2017comprehensive}. Another approach simplifies the nonlinear longitudinal dynamics of vehicles by transforming them into a linearized form to facilitate solutions to speed and charge-planning problems~\cite{KSAE}. In addition, research has focused on determining speed profiles to minimize energy consumption while driving~\cite{Sciarretta2015,guanetti2018control,Vahidi2018}. Eco-driving and eco-tracking are also closely related to supervisory control and powertrain management because they optimize vehicle fuel efficiency while promoting sustainable driving practices~\cite{Guzzella2013}.

When the driving route is predetermined, the goal of eco-driving is to determine a speed profile that minimizes the cost function, encompassing metrics such as the travel time (including charging time) and energy consumption over the entire predicted planning horizon. Three distinct approaches have been proposed to achieve this.

One approach employs eco-driving strategies using cruise control frameworks based on spatial or distance-based model predictive control (MPC)~\cite{Weimann2017ifac,Lim2017,Jia2019ifac}. In~\cite{Jia2020tvt}, a simulation environment leveraging V2X information was developed, emphasizing nonlinear vehicle dynamics. Another widely studied method is dynamic programming (DP), with various strategies proposed for solving eco-driving problems~\cite{Chao2020,Ozatay2014,kim:access:2021,kim:tits:2021,bae2022gaussian}. 

Recent advancements in computational performance have enabled the application of reinforcement learning to eco-driving MPC problems, particularly in connected and automated vehicles. Model-based reinforcement learning has shown promise for enhancing the performance of eco-driving MPC frameworks~\cite{Lee2020RL}. In addition, non-model-based reinforcement learning approaches, such as the deep deterministic policy gradient and twin delayed deep deterministic policy gradient, have been explored to enable MPC through learning-based models derived from surrounding driving environments~\cite{Zhou2020DDPG,Wegener2021TD3}.

Various methods have been proposed to optimize charging plans for EVs. In~\cite{BENZ}, DP and mixed-integer linear programming were employed to develop a charging strategy that accounts for the time spent traveling to and from charging stations along a driving route. A more practical approach to optimal charging plans was introduced in~\cite{VOLVO} by incorporating nonlinear models that considered the thermal management of battery heating and cooling. In addition,~\cite{PORSCHE} presented an optimization framework that addresses charging infrastructure, driving range, operational costs, and battery reliability to support the broader adoption of EVs. In~\cite{BMW}, a mixed-integer nonlinear programming (MINLP) approach combined speed planning with charging strategies and compared the results with simplified linear programming and DP-based methods. This approach utilizes planned route information from a separate network and involves a relatively complex model.

In this paper, we present a trip-planning method designed to minimize energy consumption and reduce travel time, including charging time, for battery-powered EVs by utilizing real-time V2X data. This method integrates speed planning with optimal charging strategies and leverages advanced optimization techniques to improve efficiency and practicality. The main contributions of this study are summarized as follows. 
\begin{itemize}  
\item
\emph{Integrated Optimization Framework:} We propose a comprehensive optimization framework that combines nonlinear programming (NLP), convex quadratic programming (QP), and mixed-integer quadratic programming (MIQP) for simultaneous speed and charging planning.  
\item
\emph{Real-Time V2X Integration:} The framework utilizes real-time V2X communication to incorporate route geometry, traffic conditions, and charging infrastructure, enabling a practical implementation for connected EVs.  
\item 
\emph{Comparative Analysis:} A thorough evaluation of different optimization approaches, including NLP, QP, MINLP, and MIQP, is performed to highlight trade-offs in computational efficiency and solution quality.  
\item
\emph{Scalability and Real-Time Capability:} The proposed MIQP-based method demonstrates exceptional computational efficiency, solving trip plans exceeding 700 km within a few seconds, making it suitable for real-world applications.  
\end{itemize}

The remainder of this paper is organized as follows. Section~\ref{sec:problem} introduces the modeling approach and formulates the optimal control problem (OCP) for the simultaneous speed and charging planner. Section~\ref{sec:method} describes the process used to enhance the proposed algorithm in detail. Section~\ref{sec:sim} provides a comparative analysis of the results obtained from NLP, QP, MINLP, and MIQP approaches and investigates the relationship between the trip time and energy consumption. Section~\ref{sec:discussion} presents a detailed discussion of the solution methodology and simulation results. Finally, Section~\ref{sec:conclusion} offers concluding remarks and outlines directions for future research.

\section{Problem Statement: EV Optimal Trip Planning}
\label{sec:problem}
\begin{figure}[t]
	\centering
	\includegraphics[width=.5\textwidth]{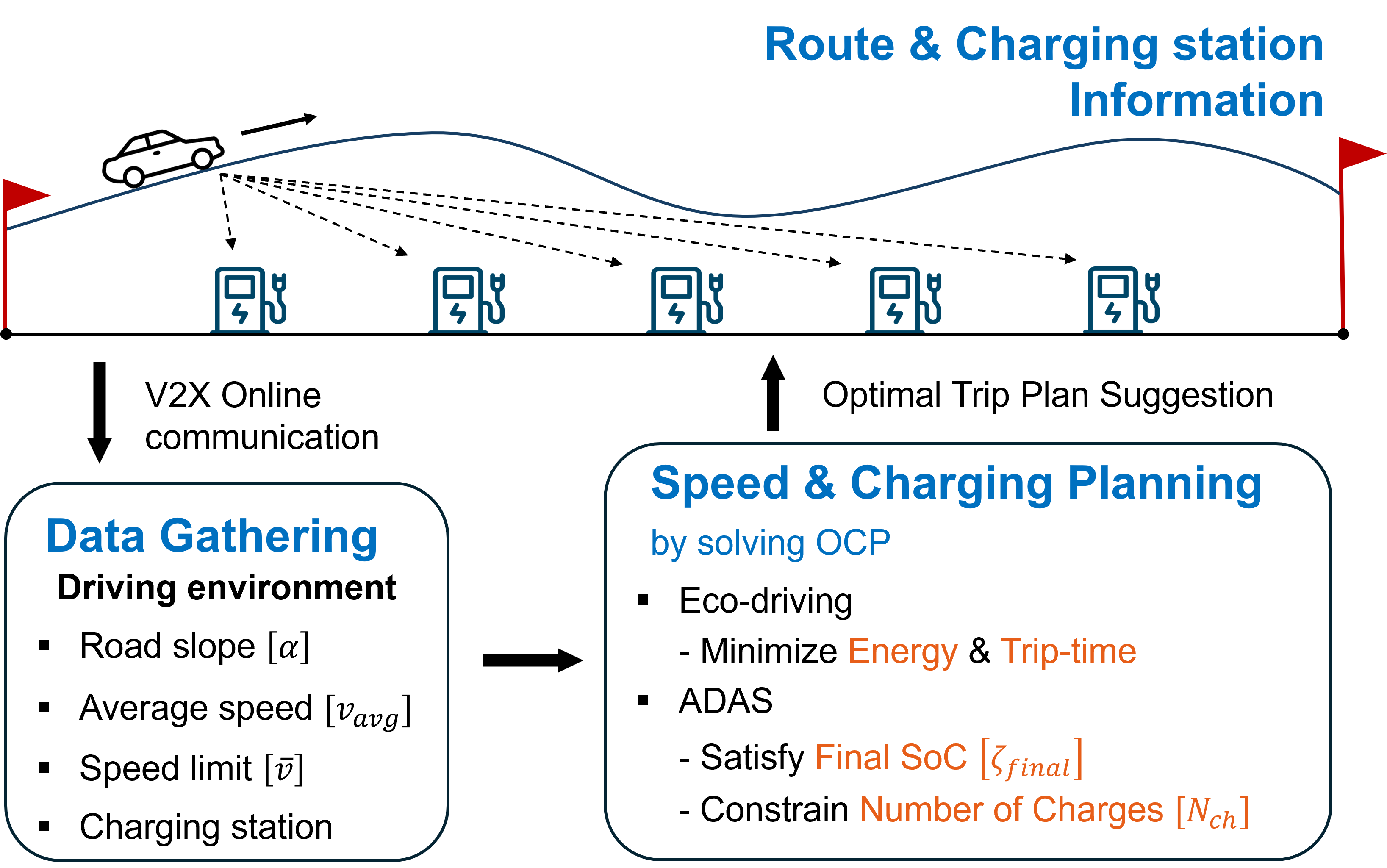}
	\caption{Conceptual diagram of EV trip (vehicle speed and charging strategy) planning using V2X connectivity.}
	\label{fig:diagram}
\end{figure}

Fig.~\ref{fig:diagram} illustrates the schematic diagram of the proposed optimization-based trip planner, which simultaneously optimizes vehicle speed and charging strategy by leveraging driving environment information obtained through V2X connectivity. Geographic road information, such as road slopes at specific spatial points, is incorporated into the vehicle's longitudinal dynamics. Traffic data, including average speed along a given route and speed limits at key spatial points, are utilized to formulate state constraints. Charging station specifications and related information are integrated into the charging strategy. The combined information available through V2X connectivity is seamlessly incorporated into the optimal control problem (OCP) of the proposed \textbf{evS$^{2}$CP}.

\begin{figure}[t]
	\centering
	\includegraphics[width=.375\textwidth]{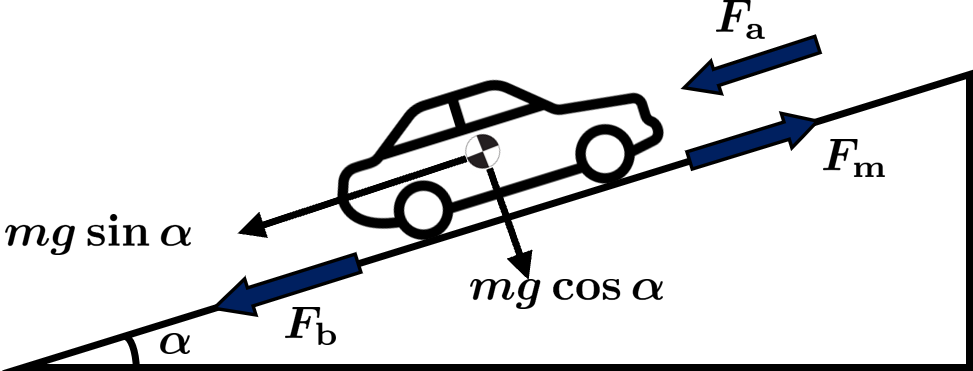}
	\caption{Schematics of the longitudinal vehicle dynamics.}
	\label{fig:dynamics}
\end{figure}

\subsection{Modeling}
\subsubsection{Vehicle dynamics}
For dynamical system modeling of EVs, we consider the following longitudinal vehicle dynamics:
\begin{align}\label{eq:EVdymamics}
	m\dot{v}(t) = F_{\rm m}(t) - F_{\rm b}(t) - F_{\rm r}(t)
\end{align}
where $v(t)$ denotes the longitudinal speed of the vehicle, $m$ is the vehicle mass, and $F_{\rm r}(t) = F_{\rm a}(t) + F_{\alpha}(t)$ is the combined load force.
The load force $F_\mathrm{air}(t)$ denotes the longitudinal aerodynamic drag. Assuming that the wind speed is relatively low, the aerodynamic drag force can be expressed as
\begin{align}\label{eq:aerodrag}
	F_{\rm a}(t) = \frac{1}{2}\rho_\mathrm{a} C_\mathrm{d} A_\mathrm{f} v^2(t)
\end{align}
where $\rho_a$ denotes the mass density of air, $C_d$ is the aerodynamic drag coefficient, and $A_f$ is the frontal area of the vehicle.
The gravitational load force $F_{\alpha}(t)$ refers to the road load
\begin{align}\label{eq:roadload}
	\displaystyle F_{\alpha}(t) = mg\!\left(\sin(\alpha(s(t)))+C_\mathrm{r}\cos(\alpha(s(t)))\right)
\end{align}
where $g$ is the gravitational acceleration, $C_\mathrm{r}$ is the rolling resistance coefficient, $\alpha(s(t))$ is the road slope, and $s(t)$ is the travel distance.

\paragraph{Discrete-time model}\label{subsec:timedynamics}
For time-discretization, we consider the Euler forward method with a sampling time interval $\Delta t_k$ for each time step $k$:
\begin{equation}\label{eq:v_euler1}
	v(t+\Delta t_k) = v(t) + \frac{F_{\rm m}(t) - F_{\rm b}(t) - F_{\rm r}(t)}{m} \Delta t_k
\end{equation}
that can be rewritten as 
\begin{equation}\label{eq:v_euler2}
	v(t_{k+1}) = v(t_k) + \frac{F_{\rm m}(t_{k}) - F_{\rm b}(t_{k}) - F_{\rm r}(t_{k})}{m} \Delta t_k \,.
\end{equation}
By defining $v_{k} :=v(t_k)$ and $s_{k} :=s(t_k)$, we can rewrite this as
\begin{equation}\label{eq:v_euler3}
	v_{k+1} = v_{k} + \frac{F_{{\rm m},k} - F_{{\rm b},k} - F_{{\rm r},k}}{m} \Delta t_k
\end{equation}
where the combined load force $F_{{\rm r},k}$ at time step $k$ is given by:
\begin{equation}\label{eq:totalload}
	F_{{\rm r},k}(v_k) = mg\!\left(\sin\alpha(s_{k}) \!+\! C_\mathrm{r}\cos\alpha(s_{k})\right) + \frac{1}{2}\rho_\mathrm{a} C_\mathrm{d} A_\mathrm{f} v_{k}^2 \,.
\end{equation}

\paragraph{Dynamics in the spatial domain}\label{subsec:spatialdynamics}
To rewrite the vehicle dynamics in the spatial domain, we substitute the time-dependent variables with their spatial equivalents, where the independent variable becomes the travel distance $s$ (or location index along the route, denoted as $s_{k}$). Assuming constant acceleration during the time interval $[t_{k}, t_{k+1}]$, we have
\begin{equation}\label{eq:delta_sk}
	\Delta s_k := s(t_{k+1})-s(t_{k}) = {v(t_{k})+v(t_{k+1}) \over 2}\Delta t_k
\end{equation}
that can be rewritten as
\begin{equation}\label{eq:delta_tk}
	\Delta t_k = {2 \Delta s_k \over v_{k} + v_{k+1}} \,.
\end{equation}
By substituting ~\eqref{eq:delta_tk} into~\eqref{eq:v_euler3}, we obtain
\begin{equation}\label{eq:v2diff}
	v_{k+1}^2 = v_{k}^2 + \frac{2 \Delta s_k}{m} \left( F_{{\rm m},k} - F_{{\rm b},k} - F_{{\rm r},k} (v_{k}^2, \alpha_{k}) \right)
\end{equation}
where $F_{{\rm r},k} (v_{k}^2, \alpha_{k})$ is affine in $v_{k}^2$. Here, we use the notation $v_{k} = v(t_{k}) = v(s(t_{k})) = v(s_{k})$, and other dependent variables follow the same notation. This spatial formulation facilitates the optimal planning of speed and energy consumption along the route and serves as the foundation for spatial domain planning strategies in electric vehicles.

\subsubsection{Traction force limit}\label{subsec:traclimit}
The traction force of the EV (that is, the motor) has physical limitations based on either the maximum mechanical torque or power generation:
\begin{equation}\label{eq:maxfm}
	\overline{F}_{\mathrm{m},k} = \min\left\{\overline{F}_{\mathrm{m}}, \overline{F}_{\mathrm{m},k}^\mathrm{cp}(v_{k})\right\}
\end{equation}
This can be rewritten as an approximation:  
\begin{equation}\label{eq:fmapprox}
\begin{split}
& 0 \leq F_{\mathrm{m},k} \leq \overline{F}_{\mathrm{m}}\,, \\
& 0 \leq F_{\mathrm{m},k} \leq \overline{F}_{\mathrm{m},k}^\mathrm{cp}(x_{k})
= \frac{\overline{P}_{\mathrm{m}}}{\sqrt{x_{k}}}
	\approx c_{0} + c_{1} x_{k}
\end{split}
\end{equation}
where $x_{k} := v_{k}^2$ is defined as the squared vehicle speed. These constraints are jointly linear in $(F_{\mathrm{m},k},x_{k})$, where $\mathrm{cp}$ is the constant power of the electric motor.

For precision and practicality, additional constraints, $ F_{\text{m},k}\cdot F_{\text{b},k}=0 $, should be considered, because we do not consider regenerative braking for which the motor force can have negative $ F_{\text{m},k}<0 $. For simplicity, these constraints are ignored because a well-defined optimization automatically satisfies these constraints.

\subsubsection{Battery dynamics}\label{subsec:battdaynamics}
The change in the battery state of charge (SoC) owing to the energy consumption of the traction motor expressed within the spatial domain is given as
\begin{equation}\label{eq:zeta_euler1}
	\zeta_{k+1}=\zeta_k-\frac{\Delta s_k}{\eta_k E_\mathrm{cap}}F_{\mathrm{m},k}
\end{equation}
where $\zeta_k$ is the battery SoC at the current location index $k$; $\Delta s_k$ is the distance traveled between indices $k$ and $k+1$; $\eta_k$ is the electrical efficiency of the motor, which depends on the vehicle's driving speed and traction force; $F_{\mathrm{m},k}$ is the traction force exerted by the motor; and $E_\mathrm{cap}$ is the battery capacity. This equation models the decrement in the battery's SoC as a function of the energy consumed by the traction motor over a discrete spatial step $\Delta s_k$. It captures the impact of motor efficiency and battery capacity on energy usage, making it suitable for optimization-based energy-management strategies in EVs.

\subsection{Optimal Speed Planner}
\subsubsection{Objectives}
We present the following multiple objectives for optimization to derive the optimal driving speed, considering both eco-driving and trip-time reduction simultaneously.
\noindent
\paragraph{Trip-time minimization}
The trip-time is
\begin{equation}\label{eq:tN}
	t_N = \sum_{k=0}^{N-1} \Delta t_{k}
	= \sum_{k=0}^{N-1} {2 \Delta s_k \over v_{k} + v_{k+1}}
	\left(\mbox{or}
	\sum_{k=0}^{N-1} {\Delta s_k \over v_{k}} \right)
	\!.
\end{equation}
For trip time minimization, we consider the following cost function: 
\begin{equation}\label{eq:J1}
	J_{1} = t_{N}
\end{equation}
or its approximation
\begin{equation}\label{eq:J1hat}
	\hat{J}_{1} = - \sum_{k=0}^{N-1} \frac{v_{k}^{2}}{\Delta s_{k}^2}
\end{equation}
that is linear in $v_{k}^{2}$. 

\subsubsection{Energy-consumption minimization}
The motor electric power for the traction mode can be represented as follows: 
\begin{equation}\label{eq:Pelec}
	P_{\mathrm{elec}} (F_{m}, v) = \frac{F_{\mathrm{m}} v}{\eta_{\rm m}(F_{\mathrm{m}}, v)}
\end{equation}
where $\eta_{\rm m}(F_{\mathrm{m}}, v)$ is the motor efficiency of the power conversion from electrical to mechanical,
and $P_{\mathrm{mech}}=F_{\mathrm{m}} v$ denotes the mechanical power. 

For energy consumption minimization over a predefined route of trip, we consider a cost function
\begin{equation}\label{eq:J2}
\begin{split}
J_{2} 
&=\! \sum_{k=0}^{N-1} P_{\mathrm{elec}} (F_{\mathrm{m},k}, v_{k}) \Delta t_{k} 
=\! \sum_{k=0}^{N-1} \frac{F_{\mathrm{m},k} v_{k}}{\eta_{\rm m}(F_{\mathrm{m}}, v)} \Delta t_{k} \\
& =\! \sum_{k=0}^{N-1} \frac{F_{\mathrm{m},k}}{\eta_{\rm m}(F_{\mathrm{m}}, v)} v_{k}\Delta t_{k} 
\approx \! \sum_{k=0}^{N-1} F_{\mathrm{elec}} (F_{\mathrm{m},k}, v_{k}) \Delta s_{k} 
\end{split}
\end{equation}
where $F_{\mathrm{elec}} (F_{\mathrm{m},k}, v_{k}) = {F_{\mathrm{m},k}}/{\eta_{\rm m}(F_{\mathrm{m}}, v)}$;
or its approximation in a quadratic form 
\begin{equation}\label{eq:J2hat}
\hat{J}_{2} = \sum_{k=0}^{N-1} \begin{bmatrix} v_{k}^2 \\ F_{\mathrm{m},k} \\ 1 \end{bmatrix}^{\!\top} \!\!
\begin{bmatrix} q_{2} & h & q_{1}  \\ h & r_{2} & r_{1} \\ q_{1} & r_{1} & h_0 \end{bmatrix} 
\begin{bmatrix} v_{k}^2 \\ F_{\mathrm{m},k} \\ 1 \end{bmatrix} 
\!\Delta_{k}
\end{equation}
where $\Delta_{k}$ is either $\Delta t_{k}$ in the time domain or $\Delta s_{k}$ in the spatial domain.
We consider a regression model that approximates $P_{\mathrm{elec}} (v^2, F_{m})$ or $F_{\mathrm{elec}} (v^2, F_{m})$:
\begin{equation}\label{eq:Pelecapprox}
	P_{\mathrm{elec}}  \mbox{ or } F_{\mathrm{elec}}  \approx 
	\begin{bmatrix} v^2 \\ F_{\rm m} \\ 1 \end{bmatrix}^{\top} 
	\begin{bmatrix} q_{2} & h & q_{1}  \\ h & r_{2} & r_{1} \\ q_{1} & r_{1} & h_0 \end{bmatrix} 
	\begin{bmatrix} v^2 \\ F_{\rm m} \\ 1 \end{bmatrix}
\end{equation}
where the cost function corresponding to the energy consumption can be rewritten as 
\begin{equation}\label{eq:J2hat2}
\hat{J}_{2} 
= \!\! \sum_{k=0}^{N-1}\! P_{\mathrm{elec}} (v_{k}^2, F_{\mathrm{m},k}) \Delta t_{k} 
\!= \!\! \sum_{k=0}^{N-1}\! F_{\mathrm{elec}} (v_{k}^2, F_{\mathrm{m},k}) \Delta s_{k}
\end{equation}
which is jointly quadratic in $(v_{k}^{2}, F_{\mathrm{m},k})$. 

For an (approximate) strongly convex program, we assume that the Hessian matrix of the cost $\hat{J}_{2}$ is positive definite.
The regression model for a motor efficiency map should depend on the domain of the problem formulation.
If the OCP is formulated in the time domain, the regression model is obtained for $P_{\mathrm{elec}} (F_{\mathrm{m},k}, v_{k})={v F_{\mathrm{m}}}/{\eta_{\rm m}(F_{\mathrm{m}}, v)}$, whereas 
the regression model of $F_{\mathrm{elec}} (F_{\mathrm{m},k}, v_{k})={F_{\mathrm{m}}}/{\eta_{\rm m}(F_{\mathrm{m}}, v)}$ is used to approximate the energy consumption in the OCP in the space domain.

\subsubsection{Additional costs}
In addition to trip time and energy consumption minimization, we also consider the minimization of the (hydraulic) mechanical braking energy using a quadratic function:
\begin{equation}\label{eq:J3hat}
\begin{split}
	\hat{J}_{3}
	& \approx \sum_{k=0}^{N-1} b_2 F_{\mathrm{b},k}^2\Delta s_k \ .
\end{split}
\end{equation}
where $\Delta_{k}$ is either $\Delta t_{k}$ in the time domain or $\Delta s_{k}$ in the spatial domain.

\subsubsection{Constraints with real-time data}\label{subsec:constraints}
To simulate road conditions, real-time parameters were configured through V2X communication using Web API data.
The gravitational load force given in~\eqref{eq:roadload} can be rewritten in the following spatial domain form:
\begin{equation}
	F_{\alpha,k}=mg\left(\sin\alpha_k+C_\mathrm{r}\cos\alpha_k\right)
\end{equation}
and the road slope $\alpha_k$ can be calculated from the road elevation $h$ obtained from V2X data using the following equation:
\begin{equation}
	\alpha_k=\arctan\left({h_{k+1}-h_k\over\Delta s_k}\right)\,.
\end{equation}

The constraints on the driving speed are as follows: 
\begin{equation}
	0 \leq v_k \leq \overline{v}_k
\end{equation}
where $\overline{v}_k$ represents the upper bound determined using the maximum speed limit information for each road segment.
In addition, to satisfy driver preferences and maintain a minimum driving speed on highways, the real-time average speeds of the surrounding vehicles were considered.
To track the average speed, we consider a weighted objective function for $w_v(v_k-v_{\mathrm{avg},k})^2$ as a soft constraint. However, to avoid the computational instability associated with the weighting parameter, we adopted the following inequality constraint as a hard constraint:
\begin{equation}\label{eq:vavg_const}
	v_{\text{avg},k}-\sigma\leq v_k\leq v_{\text{avg},k}+\sigma
\end{equation}
where $v_{\mathrm{avg},k}$ is the average speed on the road obtained from V2X real-time data, and $\sigma$ denotes the margin that can be either a problem parameter or a slack variable. 

\subsection{Optimal Charging Planner}
We define a constraint on the destination SoC, \(\zeta_N\), based on the energy required along the route:
\begin{equation}\label{eq:zetafinal}
	\zeta_\mathrm{final} \leq \zeta_N + \delta,
\end{equation}
where \(\zeta_\mathrm{final}\) is a user-defined desired final SoC and \(\delta\) is a slack variable that provides flexibility in meeting this condition. Incorporating this inequality into the optimization problem ensures that the desired battery level is achieved at the destination. Charging events during a trip are necessary to satisfy the terminal constraints.

The total trip time, including both the driving and charging times, is expressed as
\begin{align}\label{eq:tN2}
	t_N & =\! \sum_{k=0}^{N-1}\!\!\Delta t_k + \!\!\! \sum_{k \in \mathcal{K}_\mathrm{ch}} \!\!\! \tau_{\text{ch},k} 
	= \!\sum_{k=0}^{N-1} \! \frac{2\Delta s_k}{v_k + v_{k+1}} + \!\! \sum_{k \in \mathcal{K}_{\textrm{ch}}}\!\! \tau_{\text{ch},k} \,,
\end{align}
where \(\mathcal{K}_{\textrm{ch}}\) is the set of indices corresponding to the charging station locations and \(\tau_{\text{ch},k}\) represents the charging time at each station.

To account for both driving efficiency and charging time in the optimization, the cost function from~\eqref{eq:J1hat} is updated as follows:
\begin{equation}\label{eq:J1hat2}
	\hat{J}_1 = -\sum_{k=0}^{N-1} w_\mathrm{t} \frac{v_k^2}{\Delta s_k^2} + \sum_{k \in \mathcal{K}_{\textrm{ch}}} w_\tau \tau_{\text{ch},k} \,,
\end{equation}
where \(w_\mathrm{t}\) is the weight associated with the driving speed and \(w_\tau\) is the weight associated with the charging time. The first term encourages efficient driving profiles, whereas the second term penalizes extended charging durations.

Finally, the charging time at each station is constrained by the charger specifications as follows:
\begin{equation}\label{eq:tau_const}
	0 \leq \tau_{\text{ch},k} \leq \overline{\tau}_\mathrm{ch},
\end{equation}
where \(\overline{\tau}_\mathrm{ch}\) is the maximum allowable charging time for each station.

This framework integrates the terminal SoC constraint, trip time, and charger specifications into a unified optimization problem, balancing driving efficiency and charging requirements to ensure practical and effective trip planning.

\subsection{Simultaneous Speed and Charging Planner (S$^{2}$CP)}
First, the speed planner is designed to determine the optimal speed profile that minimizes driving time. Second, the charging planner is seamlessly integrated with the speed planner to optimize the charging times at each station while ensuring that the final SoC requirement is met. This integrated approach minimizes driving energy consumption and incorporates driver convenience features within the advanced driver assistance system (ADAS).

To formulate the optimal control problem that jointly develops the speed and charging plans, the battery dynamics are updated to include the charging term in \eqref{eq:zeta_euler1} as follows:
\begin{equation}\label{eq:zeta_euler2}
	\zeta_{k+1} = \zeta_k - \frac{\Delta s_k}{\eta_k E_\text{cap}} F_{\text{m},k} + \frac{P_{\text{ch},k}}{E_\text{cap}} \tau_{\text{ch},k},
\end{equation}
where \( P_{\text{ch},k} \) denotes charging power at each station.
Additionally, an integrated cost function was formulated to optimize both the driving energy and total trip time.
\begin{equation}\label{eq:Jtotalhat}
	\hat{J} = w_\mathrm{t} t_N + \sum_{k=0}^{N-1} \left( w_\mathrm{e} F_{\text{m},k}^2 + w_\mathrm{b} F_{\text{b},k}^2 \right),
\end{equation}
where \( w_\mathrm{t} \), \( w_\mathrm{e} \), and \( w_\mathrm{b} \) are the weighting parameters representing the importance of trip time, energy consumption related to the traction force, and braking force, respectively. The traction and braking forces were modeled in quadratic form to mitigate the effects of sudden acceleration or braking and ensure smoother driving behavior.

\section{Solution Methods}\label{sec:method}
This section presents the OCPs for \textit{S$^{2}$CP} based on the issues highlighted in Section~\ref{sec:problem} and introduces the mixed-integer problem for charging instances in convexified form.

\subsection{Convex QP for Speed Planner}\label{subsec:speedplanner}
The speed planner predicts the energy consumption of the EV while driving and concurrently establishes an optimal speed plan to minimize it.

\subsubsection{NLP for Speed Planner}
For eco-driving, we define the following nonlinear OCP exclusively for the speed planner based on the approach outlined in \cite{FORCESPRO}, excluding the terms related to the charging strategy:
\begin{equation}\label{eq:nlp}
	\begin{split}
		\min \ & w_\mathrm{t} t_N + \sum_{k=0}^{N-1} \left(w_\mathrm{e} F_{\mathrm{m},k}^2 + w_\mathrm{b} F_{\mathrm{b},k}^2 \right) \,,\\
		\text{s.t.} \ & v_{k+1} = \sqrt{\frac{2\Delta s_{k}}{m_\mathrm{eq}}\left(F_{\mathrm{m},k} - F_{\mathrm{b},k} - F_{\mathrm{r},k}(v_k)\right) + v_{k}^2} \,, \\
		& t_{k+1} = t_{k} + \frac{\Delta s_{k}}{v_{k}} , \ \zeta_{k+1} = \zeta_{k} - \frac{\Delta s_{k}}{\eta_{k} E_\mathrm{cap}} F_{\mathrm{m},k} \,,\\
		& \underline{v} \leq v_{k} \leq \overline{v}_{k}, \, \underline{\zeta} \leq \zeta_{k} \leq \bar{\zeta} \,, v_0 = v_\mathrm{init}, \ \zeta_0 = \zeta_\mathrm{init} \\
		& v_{\mathrm{avg},k} - \sigma \leq v_k \leq v_{\mathrm{avg},k} + \sigma \,,\\
		& 0 \leq F_{\mathrm{m},k} \leq \overline{F}_{\mathrm{m},k}, \ 0 \leq F_{\mathrm{b},k} \leq \overline{F}_\mathrm{b} \,. \\
	\end{split}
\end{equation}

\paragraph{Motor efficiency and traction force constraints}\label{para:conditions_nlp}
The motor efficiency \( \eta_k \) depends on both the state and input variables.
\begin{equation}\label{eq:eta_nlp}
	\eta_k = \eta_k\left(F_{\mathrm{m},k}, v_k\right).
\end{equation}
The traction force constraints are defined as shown in~\eqref{eq:maxfm}, in conjunction with the force corresponding to constant power:
\begin{equation}\label{eq:maxfm2}
	\overline{F}_{\mathrm{m},k} = \min \left\{\overline{F}_\mathrm{m}, \overline{F}_{\mathrm{m},k}^\mathrm{cp}(v_k)\right\}, \quad \overline{F}_{\mathrm{m},k}^\mathrm{cp}(v_k) = \frac{\overline{P}_\mathrm{m}}{v_k},
\end{equation}
where \( \overline{F}_\mathrm{m} \) and \( \overline{P}_\mathrm{m} \) denote the maximum motor force and power limit, respectively.

\paragraph{State and control variables}\label{para:variables_nlp}
The problem uses the following variables: state variables \( X := [v, t, \zeta]^\top \) and control variables \( U := [F_{\mathrm{m}}, F_{\mathrm{b}}]^\top \).

This formulation provides a structured approach for deriving optimal speed profiles that minimize energy consumption and driving time while adhering to motor and vehicle dynamics.

\subsubsection{Convexification by approximation}
For the convex optimal control associated with eco-driving, we considered the following convex QP:
\begin{equation}\label{eq:qp}
	\begin{split}
		\min \ & -\sum_{k=0}^{N-1}{w_\mathrm{t} {x_k \over \Delta s_k^2} } + \sum_{k=0}^{N-1} \left(w_\mathrm{e} F_{\mathrm{m},k}^2 + w_\mathrm{b} F_{\mathrm{b},k}^2 \right) \,, \\
		\text{s.t.} \ & x_{k+1} = {\frac{2\Delta s_{k}}{m_\mathrm{eq}}\left(F_{\mathrm{m},k} - F_{\mathrm{b},k} - F_{\mathrm{r},k}(x_k)\right) + x_{k}} \,, \\
		& \zeta_{k+1} = \zeta_{k} - \frac{\Delta s_{k}}{\eta E_\mathrm{cap}}F_{\mathrm{m},k} \,, \\ 
		& \underline{v}^2 \leq x_{k} \leq \overline{v}_{k}^2, \, \underline{\zeta} \leq \zeta_{k} \leq \bar{\zeta} \,, x_0 = v^2_\mathrm{init}, \ \zeta_0 = \zeta_\mathrm{init} \\
		& (v_{\mathrm{avg},k}-\sigma)^2 \leq x_k \leq (v_{\mathrm{avg},k}+\sigma)^2 \,, \\
		& 0 \leq F_{\mathrm{m},k} \leq \overline{F}_{\mathrm{m},k} , \ 0 \leq F_{\mathrm{b},k} \leq \overline{F}_\mathrm{b} \,, \\
		\end{split}
\end{equation}
where $ x_k := v_k^2 $ is defined as the squared vehicle speed.

\paragraph{Motor efficiency and traction force constraints}\label{para:conditions_qp}
The motor efficiency is approximated as a constant, $\eta := \eta_k$ for all $k$. The motor force limit $ \overline{F}_{\mathrm{m},k}(x_k) $ is defined as the ones given in~\eqref{eq:fmapprox}.

\paragraph{Resistive force in vehicle dynamics}\label{para:loadforce_qp}
The resistive force owing to air drag and road geometry is defined as
\begin{equation}\label{eq:loadaffine}
	F_{\mathrm{r},k}(x_k) = mg\!\left(\sin{\alpha_k} + C_\mathrm{r}\cos{\alpha_k}\right) + {1\over2}\rho_\mathrm{a}C_\mathrm{d}A_\mathrm{f}x_k \ .
\end{equation}
This equation derived from~\eqref{eq:totalload}, is affine with respect to $x_k $, leading to the dynamics as an equality constraint in optimization-based planning.

\paragraph{State and control variables}\label{para:variables_qp}
The QP problem uses the following variables: state variables \( X := [x, \zeta]^\top \) and control variables \( U := [F_{\mathrm{m}}, F_{\mathrm{b}}]^\top \). The state variable $t$ which is part of the NLP~\eqref{eq:nlp}, is excluded here to linearize the dynamics.

\subsection{Mixed-Integer QP for Simultaneous Speed and Charging Planner}\label{subsec:method_mip}
We now consider a system that uses a charging planner to formulate an optimal charging strategy based on the minimum energy consumption of an EV, which is calculated using the speed planner designed in Section~\ref{subsec:speedplanner}. Accordingly, the following discussion sequentially addresses the charging amount and number of charging events considering the Mixed-Integer Programming (MIP) problem.
\subsubsection{NLP for S$^{2}$CP}\label{subsubsec:nlp_s2cp}
From \eqref{eq:nlp}, we can update the following OCP which includes both speed and charging planning:
\begin{equation}\label{eq:nlp_s2cp}
	\begin{split}
		\min \ & w_\mathrm{t} t_N + w_d\delta + \sum_{k=0}^{N-1} \left(w_\mathrm{e} F_{\mathrm{m},k}^2 + w_\mathrm{b} F_{\mathrm{b},k}^2 \right) \,, \\
		\text{s.t.} \, \ & v_{k+1} = \sqrt{\frac{2\Delta s_{k}}{m_\mathrm{eq}}\left(F_{\mathrm{m},k} - F_{\mathrm{b},k} - F_{\mathrm{r},k}(v_k)\right) + v_{k}^2} \,, \\
		& t_{k+1} = t_{k} + {\Delta s_{k} \over v_{k}} + \tau_{\mathrm{ch},k} \,, \\
		&\zeta_{k+1} = \zeta_{k} - \frac{\Delta s_{k}}{\eta_{k} E_\mathrm{cap}}F_{\mathrm{m},k} + \frac{P_{\mathrm{ch},k}}{E_\mathrm{cap}}\left(\tau_{\mathrm{ch},k}-\underline{\tau}_\mathrm{ch}\right) \,, \\
		& \underline{v} \leq v_{k} \leq \overline{v}_{k}, \, \underline{\zeta} \leq \zeta_{k} \leq \bar{\zeta} \,, v_0 = v_\mathrm{init}, \ \zeta_0 = \zeta_\mathrm{init} \\
		& \underline{\tau}_\mathrm{ch} \leq \tau_{\mathrm{ch},k} \leq \overline{\tau}_\mathrm{ch} \,, v_{\mathrm{avg},k}-\sigma \leq v_k \leq v_{\mathrm{avg},k}+\sigma , \\
		& 0 \leq F_{\mathrm{m},k} \leq \overline{F}_{\mathrm{m},k} , \ 0 \leq F_{\mathrm{b},k} \leq \overline{F}_\mathrm{b} \,, \\
		& \zeta_\mathrm{final} \leq \zeta_N + \delta , \ 0 \leq \delta \,, \\
	\end{split}
\end{equation}
where $ \underline{\tau}_\mathrm{ch} $ is the waiting time in a queue that must be spent at station $ k $ for charging.

\paragraph{Motor efficiency, traction force, and charging power}\label{para:conditions_nlp_s2cp}
The motor efficiency $\eta_k$ and traction force limit $\overline{F}_{\mathrm{m},k}$ are defined as described in Section~\ref{para:conditions_nlp}. In this NLP problem, the charging power is defined as a profile with respect to the battery SoC as $P_{\mathrm{ch},k} := P_\mathrm{ch}(\zeta_k)$.

\paragraph{Charging amount corresponding to the final SoC requirement}\label{para:chargingamount}
In this OCP, ~\eqref{eq:zetafinal} is implemented to derive the optimal charging amount required to satisfy the final SoC. 
As shown in~\eqref{eq:nlp_s2cp}, the effective charging time $\tau_{\mathrm{ch},k}-\underline{\tau_\mathrm{ch}}$ excluding the waiting time is incorporated to reflect the increase in battery energy. 
In addition, the total trip time including the time spent at the charging station, is incorporated into the cost function as $w_t t_N$.

\paragraph{State and control variables}\label{para:variables_nlp_s2cp}
The NLP problem considering charge planning uses the following variables: state variables \( X := [v, t, \zeta]^\top \) and control variables \( U := [F_{\mathrm{m}}, F_{\mathrm{b}}, \tau_{\mathrm{ch}}, \delta]^\top \), which include the charging time at each station and the slack variable for the final SoC.

\subsubsection{QP for S$^{2}$CP}\label{subsubsec:qp_s2cp}
Similar to the approach presented inSection~\ref{subsubsec:nlp_s2cp}, we consider the following OCP that integrates charging planning into the convex QP problems ~\eqref{eq:qp}:
\begin{equation}\label{eq:qp_s2cp}
	\begin{split}
		\min \ & -\sum_{k=0}^{N-1}{w_\mathrm{t} {x_k \over \Delta s_k^2} } + \sum_{k\in\mathcal{K}_{\textrm{ch}}}w_\tau \tau_{\mathrm{ch},k} \\
		&+ w_d\delta + \sum_{k=0}^{N-1} \left(w_\mathrm{e} F_{\mathrm{m},k}^2 + w_\mathrm{b} F_{\mathrm{b},k}^2 \right) \,, \\
		\text{s.t.} \ \ & x_{k+1} = {\frac{2\Delta s_{k}}{m_\mathrm{eq}}\left(F_{\mathrm{m},k} - F_{\mathrm{b},k} - F_{\mathrm{r},k}(x_k)\right) + x_{k}} \,, \\
		& \zeta_{k+1} = \zeta_{k} - \frac{\Delta s_{k}}{\eta E_\mathrm{cap}}F_{\mathrm{m},k} + \frac{P_\mathrm{ch}}{E_\mathrm{cap}}\left(\tau_{\mathrm{ch},k}-\underline{\tau}_\mathrm{ch}\right) \,, \\
		& \underline{v}^2 \leq x_{k} \leq \overline{v}_{k}^2 ,\,  \underline{\zeta} \leq \zeta_{k} \leq \bar{\zeta} ,\,  x_0 = v^2_\mathrm{init} , \,\zeta_0 = \zeta_\mathrm{init} , \\
		& \underline{\tau}_\mathrm{ch} \leq \tau_{\mathrm{ch},k} \leq \overline{\tau}_\mathrm{ch} \\
		& (v_{\mathrm{avg},k}-\sigma)^2 \leq x_k \leq (v_{\mathrm{avg},k}+\sigma)^2 \\
		& 0 \leq F_{\mathrm{m},k} \leq \overline{F}_{\mathrm{m},k} , \ 0 \leq F_{\mathrm{b},k} \leq \overline{F}_\mathrm{b} \,, \\
		& \zeta_\mathrm{final} \leq \zeta_N + \delta, \ 0 \leq \delta \,. \\
	\end{split}
\end{equation}

\paragraph{Motor efficiency, traction force, and charging power}\label{para:conditions_qp_s2cp}
The motor efficiency $\eta$ and charging power $P_\mathrm{ch}$ are approximated as constants to convexify the model dynamics through linearization, $\eta:=\eta_k$ and $P_\mathrm{ch}:=P_{\mathrm{ch},k}$ for all $k$.
Additionally, traction force limit is represented as shown in~\eqref{eq:fmapprox}.

\paragraph{Weight separation for total trip time}\label{para:weightseperation}
The objective terms of the net driving time $-w_\mathrm{t}{x_k\over\Delta s_k^2}$ and charging time $w_\tau \tau_{\mathrm{ch},k}$ are decoupled from the objective of trip time $t_N$ in the NLP-based OCP~\eqref{eq:nlp_s2cp}, and different values can be assigned to the weighting parameters $w_\mathrm{t}$ and $w_\tau$.

\paragraph{State and control variables}\label{para:variables_qp_s2cp}
The QP problem considering charge planning uses the following variables: state variables \( X := [x, \zeta]^\top \) and control variables \( U := [F_{\mathrm{m}}, F_{\mathrm{b}}, \tau_{\mathrm{ch}}, \delta]^\top \), which include the charging time $\tau_\mathrm{ch}$ at each station and slack variable $\delta$ for final SoC. The state variable $t$ which is part of the NLP~\eqref{eq:nlp_s2cp}, is excluded here to linearize the dynamics, whereas it is considered only in the cost function.

\subsubsection{MINLP for S$^{2}$CP}\label{subsubsec:minlp}
We now consider a mixed-integer program along with the problem of station selection by reformulating NLP~\eqref{eq:nlp_s2cp}. To model the associated discrete constraints on the number of charges, we define the indicator variables $ z_k \in \{0, 1\} $, where $ z_k=0 $ indicate non-charging at station $ k $ when $ \tau_{\mathrm{ch},k}=0 $, and $ z_k=1 $ when $ \tau_{\mathrm{ch},k}\geq\underline{\tau}_\mathrm{ch} $. The constraint ~\eqref{eq:tau_const} can be replaced by
\begin{equation}\label{eq:tau_const2}
	z_k \underline{\tau}_\mathrm{ch} \leq \tau_{\mathrm{ch},k} \leq z_k \overline{\tau}_\mathrm{ch}\ \text{and}\ z_k \in \{0, 1\}\ .
\end{equation}
The waiting time is treated as a soft constraint in the cost function derived by substituting~\eqref{eq:tN2} into~\eqref{eq:Jtotalhat}, thereby reducing the total number of charges. Furthermore, to enforce hard constraints on the number of charges, we consider the inequality
\begin{equation}\label{eq:Nch}
	\sum_{k\in \mathcal{K}_{\textrm{ch}}}z_k \leq N_{\mathrm{ch}} \leq |\mathcal{K}_{\textrm{ch}}|\ .
\end{equation}
In addition, the charging-related battery dynamics in ~\eqref{eq:zeta_euler2} can be rewritten as
\begin{equation}\label{eq:zeta_euler3}
	\zeta_{k+1} = \zeta_k - {\Delta s_k \over \eta_k E_\mathrm{cap}}F_{\mathrm{m},k} + {P_{\mathrm{ch},k} \over E_\mathrm{cap}}(\tau_{\mathrm{ch},k}-z_k\underline{\tau}_\mathrm{ch})\ .
\end{equation}

By integrating the constraints on the number of charges and modeling the waiting time at charging stations, we consider the following OCP derived from~\eqref{eq:nlp_s2cp}:
\begin{equation}\label{eq:minlp}
		\begin{split}
			\min \ & w_\mathrm{t} t_N + w_\mathrm{d}\delta + \sum_{k=0}^{N-1} \left(w_\mathrm{e} F_{\mathrm{m},k}^2 + w_\mathrm{b} F_{\mathrm{b},k}^2 \right) \,, \\
			\text{s.t.} \ \  &\! v_{k+1} = \sqrt{\frac{2\Delta s_{k}}{m_\mathrm{eq}}\left(F_{\mathrm{m},k} - F_{\mathrm{b},k} - F_{\mathrm{r},k}(v_k)\right) + v_{k}^2} \,, \\
			&\! t_{k+1} \!= t_{k} + {\Delta s_{k} \over v_{k}} + \tau_{\mathrm{ch},k} \,, \\
			&\! \zeta_{k+1} \!= \zeta_{k} - \frac{\Delta s_{k}}{\eta_{k} E_\mathrm{cap}}F_{\mathrm{m},k} + {P_\mathrm{ch,k} \over E_\mathrm{cap}}(\tau_\mathrm{ch,k}-z_k\underline{\tau}_\mathrm{ch}) , \\
			&\! \underline{v} \leq v_{k} \leq \overline{v}_{k} , \ \underline{\zeta} \leq \zeta_{k} \leq \bar{\zeta} \,, v_0 = v_\mathrm{init} , \ \zeta_0 = \zeta_\mathrm{init} \,, \\
			&\! z_k\underline{\tau}_\mathrm{ch} \leq \tau_\mathrm{ch} \leq z_k\overline{\tau}_\mathrm{ch},\ z_k\in\{0,1\}\ \text{for}\ k\in\mathcal{K}_{\textrm{ch}} \,, \\
			&\! \sum_{k\in\mathcal{K}_{\textrm{ch}}}z_k \leq N_\mathrm{ch} \,, \\
			&\! v_{\mathrm{avg},k}-\sigma \leq v_k \leq v_{\mathrm{avg},k}+\sigma \,, \\
			&\! 0 \leq F_{\mathrm{m},k} \leq \overline{F}_{\mathrm{m},k} , \ 0 \leq F_{\mathrm{b},k} \leq \overline{F}_\mathrm{b} \,, \\
			&\! \zeta_\mathrm{final} \leq \zeta_N + \delta , \ 0 \leq \delta \,. \\
		\end{split}
\end{equation}

\paragraph{Motor efficiency, traction force, and charging power}\label{para:conditions_minlp}
The motor efficiency $\eta_k$, traction force limit $\overline{F}_{\mathrm{m},k}$ and charging power $P_{\mathrm{ch},k}$ are defined in Section~\ref{para:conditions_nlp_s2cp}.

\paragraph{Number of charge constraint}\label{para:numofcharges_minlp}
From the above OCP, $N_\mathrm{ch}$ is either the number of charges that the driver wants or a pre-defined parameter affected by the traffic environment. In this study, $ N_\mathrm{ch} $ was determined as the minimum required number of charges predicted based on the maximum battery energy consumption, assuming that the EV maintained its maximum speed. The force corresponding to the maximum speed limit on the road is derived from~\eqref{eq:totalload} and ~\eqref{eq:v2diff} as follows:
\begin{subequations}\label{eq:maxforce}
	\begin{align}
		F_{\mathrm{r},k}(\overline{v}_k) &= mg\!\left(\sin{\alpha_k} + C_\mathrm{r}\cos{\alpha_k}\right)  + {1 \over 2}\rho_\mathrm{a}C_\mathrm{d}A_\mathrm{f} \overline{v}_k^2 \\
		F_{\mathrm{m},k}(\overline{v}_k) &= {m(\overline{v}_{k+1}^2-\overline{v}_k^2) \over 2\Delta{s_k}} + F_{\mathrm{r},k}(\overline{v}_k) 
	\end{align}
\end{subequations}
where the mechanical braking force $F_{\mathrm{b},k}$ is neglected. Second, the energy consumption generated by the traction force is represented as
\begin{subequations}\label{eq:maxenergycons}
	\begin{align}
	\Delta\overline{\zeta}_\mathrm{cons} &= \sum_{k=0}^{N-1} {\Delta{s_k} \over \eta_k E_\mathrm{cap}}F_{\mathrm{m},k}(\overline{v}_k) \\
	\Delta\underline{\zeta}_\mathrm{ch,req} &= \zeta_\mathrm{final} - \zeta_\mathrm{init} + \Delta\overline{\zeta}_\mathrm{cons}
	\end{align}
\end{subequations}
where $ \Delta\overline{\zeta}_\mathrm{cons} $ and $ \Delta\underline{\zeta}_\mathrm{ch,req} $ represent the maximum possible driving energy consumption and the corresponding minimum required charging energy, respectively. Consequently, the number of charges is defined as
\begin{equation}\label{eq:minNch}
	\begin{split}
	N_\mathrm{ch} := \underline{N}_{\mathrm{ch,req}} &= \left\lceil{(1+\mu)\Delta\underline{\zeta}_\mathrm{ch,req} \over \overline{\zeta}-\underline{\zeta}}\right\rceil
	\end{split}
\end{equation}
where $ \underline{N}_\mathrm{ch,req} $ is the minimum required number of charges and $ \mu $ denotes the margin for the charging amount, which is set as a constant of 0.15.

\paragraph{State and control variables}\label{para:variables_minlp}
The MINLP problem considering charge planning uses the following variables: state variables \( X := [v, t, \zeta]^\top \) and control variables \( U := [F_{\mathrm{m}}, F_{\mathrm{b}}, \tau_{\mathrm{ch}}, z, \delta]^\top \), where the indicator variable $z$ is newly added compared to that in Section ~. ~\ref{para:variables_nlp_s2cp}.

\subsubsection{MIQP for S$^{2}$CP}
Finally, for the convex OCP ~\eqref{eq:qp_s2cp}, the optimization problem is reformulated by considering the selection of stations for charging as follows:
\begin{equation}\label{eq:miqp}
	\begin{split}
		\min \  &-\sum_{k=0}^{N-1}w_\mathrm{t}{x_k \over \Delta s_k^2} + \sum_{k\in\mathcal{K}_{\textrm{ch}}}w_{\tau}\tau_{\mathrm{ch},k} \,, \\
		& + w_\mathrm{d}\delta +  \sum_{k=0}^{N-1}\left(w_\mathrm{e} F_{\mathrm{m},k}^2 + w_\mathrm{b} F_{\mathrm{b},k}^2\right) \,, \\
		\text{s.t} \ \, &\! x_{k+1}\!={2\Delta s_k \over m_\mathrm{eq}}\left(F_{\mathrm{m},k}-F_{\mathrm{b},k}-F_{\mathrm{r},k}(x_k)\right)+x_k \,, \\
		&\! \zeta_{k+1} \!= \zeta_k - {\Delta s_k \over \eta E_\mathrm{cap}}F_{\mathrm{m},k} + {P_\mathrm{ch} \over E_\mathrm{cap}}\left(\tau_{\mathrm{ch},k}-z_k\underline{\tau}_\mathrm{ch}\right) \,, \\
		&\! \underline{v}^2 \leq x_{k} \leq \overline{v}_{k}^2 ,\, \underline{\zeta} \leq \zeta_{k} \leq \bar{\zeta} ,\,  x_0 = v^2_\mathrm{init} ,\,\zeta_0 = \zeta_\mathrm{init} , \\
		&\! z_k\underline{\tau}_\mathrm{ch} \leq \tau_{\mathrm{ch},k} \leq z_k\overline{\tau}_\mathrm{ch},\ z_k\in\{0,1\}\ \text{for}\ k\in\mathcal{K}_{\mathrm{ch}} \,, \\
		&\! \sum_{k\in\mathcal{K}_{\mathrm{ch}}}z_k \leq N_\mathrm{ch} \,, \\
		&\! (v_{\mathrm{avg},k}-\sigma)^2 \leq x_k \leq (v_{\mathrm{avg},k}+\sigma)^2 \,, \\
		&\! 0 \leq F_{\mathrm{m},k} \leq \overline{F}_{\mathrm{m},k} , \ 0 \leq F_{\mathrm{b},k} \leq \overline{F}_\mathrm{b} \,, \\
		&\! \zeta_\mathrm{final} \leq \zeta_N + \delta, \ 0 \leq \delta \,. 
	\end{split}
\end{equation}

\paragraph{Motor efficiency, traction force limit, and charging power}
The motor efficiency $\eta$, traction force limit $\overline{F}_{\mathrm{m},k}$, and charging power $P_\mathrm{ch}$ are defined in Section~\ref{para:conditions_qp_s2cp}.

\paragraph{Number of charges constraint}\label{para:numofcharges_miqp}
To consider the charging station selection problem, the number of charge limits $N_\mathrm{ch}$ is introduced in the same manner as described in Section~\ref{para:numofcharges_minlp}.

\paragraph{State and control variables}\label{para:variables_miqp}
The MINLP problem considering charging planning uses the following variables: state variables \( X := [x, \zeta]^\top \) and control variables: \( U := [F_{\mathrm{m}}, F_{\mathrm{b}}, \tau_{\mathrm{ch}}, z, \delta]^\top \) where the indicator variable $z$ is newly added compared to that in Section~\ref{para:variables_qp_s2cp}.

\section{Simulation Results}\label{sec:sim}

\begin{figure*}[t]
	\begin{subfigure}{.32\linewidth}
		\caption{\textbf{Case I}\begin{tabular}{rl}
				&Distance: 210 km \\
				&Number of stations: 5
		\end{tabular}}
		\centering
		\includegraphics[width=\textwidth]{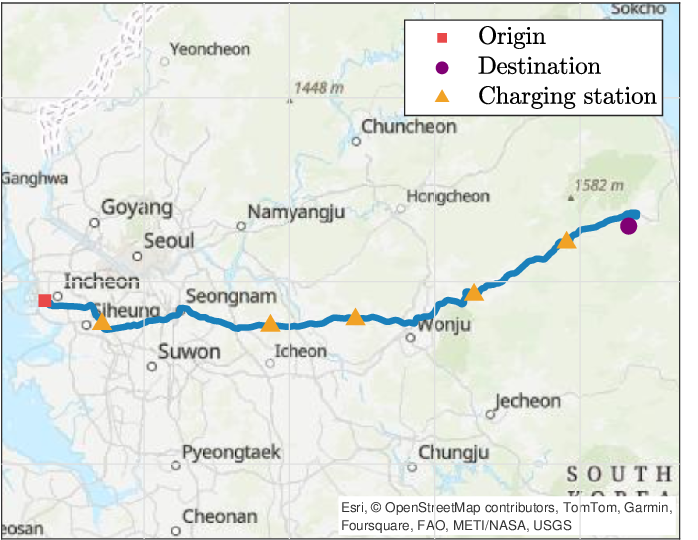}
		\label{fig:tripinfo_trip1}
	\end{subfigure}
	\hfill
	\begin{subfigure}{.32\linewidth}
		\caption{\textbf{Case II}\begin{tabular}{rl}
				&Distance: 427 km \\
				&Number of stations: 12
		\end{tabular}}
		\centering
		\includegraphics[width=\textwidth]{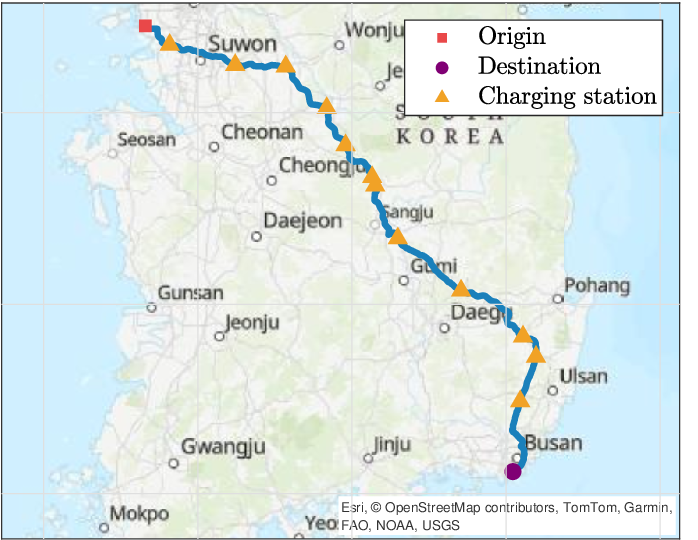}
		\label{fig:tripinfo_trip2}
	\end{subfigure}
	\hfill
	\begin{subfigure}{.32\linewidth}
		\caption{\textbf{Case III}\begin{tabular}{rl}
				&Distance: 713 km \\
				&Number of stations: 19
		\end{tabular}}
		\centering
		\includegraphics[width=\textwidth]{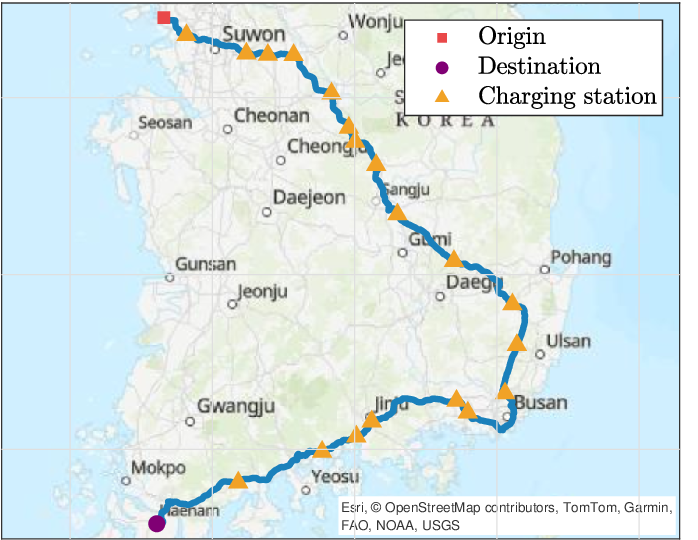}
		\label{fig:tripinfo_trip3}
	\end{subfigure}	
	\vfill
	\vspace{-2mm}
	\begin{subfigure}{.32\linewidth}
		\centering
		\includegraphics[width=\textwidth]{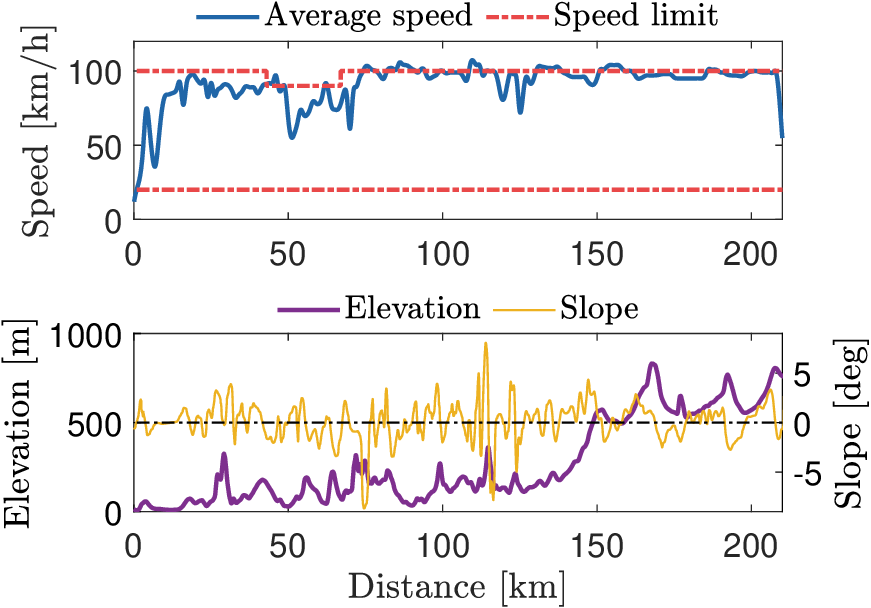}
	\end{subfigure}
	\hfill
	\begin{subfigure}{.32\linewidth}
		\centering
		\includegraphics[width=\textwidth]{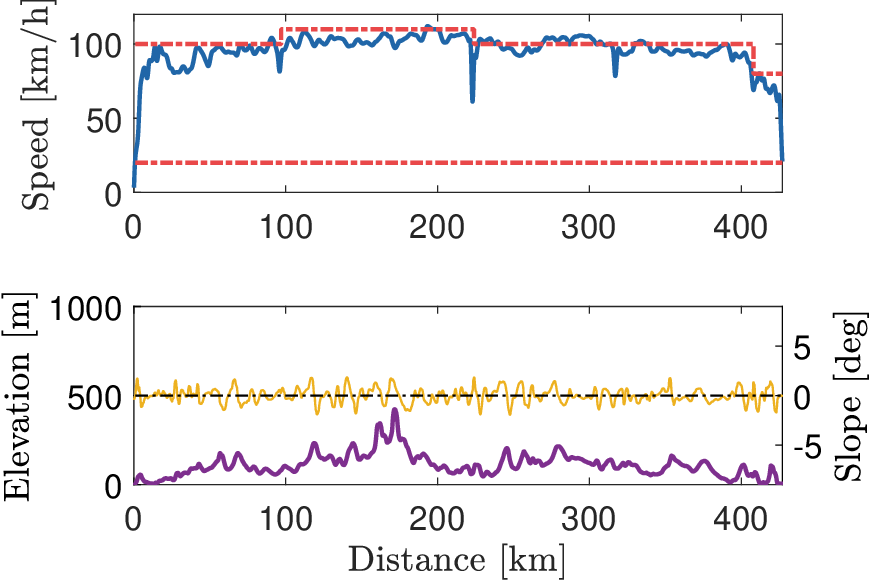}
	\end{subfigure}
	\hfill
	\begin{subfigure}{.32\linewidth}
		\centering
		\includegraphics[width=\textwidth]{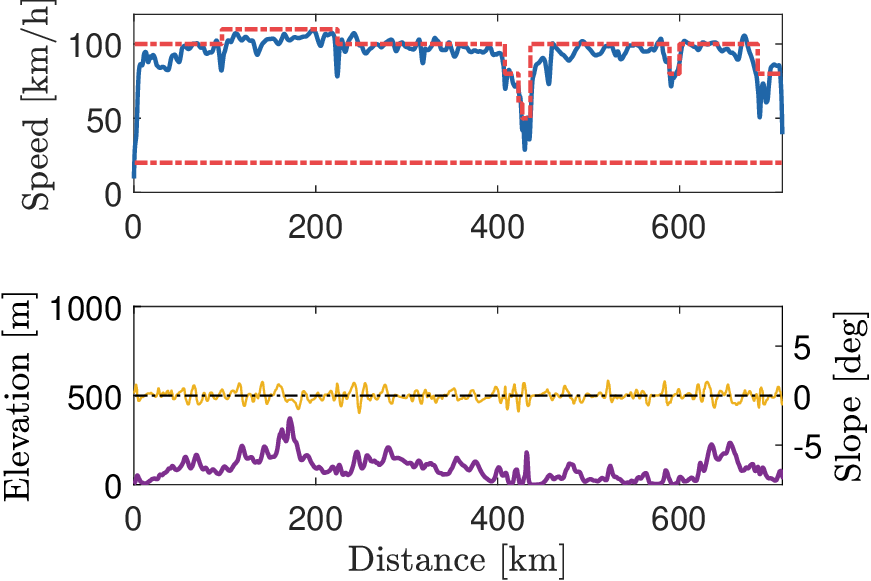}
	\end{subfigure}
	\caption{Real-time data acquisition and visualization framework for route geometry, traffic conditions, and geographic information across three driving scenarios:(a) short-distance trips ($\leq 250\:\textrm{km}$) with frequent urban and suburban stops, (b) mid-distance trips ($250\sim500\:\textrm{km}$) emphasizing intercity travel with balanced energy management, and (c) long-distance trips ($\geq 500\:\textrm{km}$) focusing on efficient charging strategies and minimal travel interruptions.}
	\label{fig:tripinfo}
\end{figure*}

\begin{table}[b!]
	\captionsetup{justification=centering}
	\caption{Vehicle Parameters for Simulation}\vspace{-3mm}\label{table:vehicle}
	\begin{center}
		{\small
			\begin{tabular}[b]{@{\,\,\,}c@{\,\,\,}@{\,\,\,}c@{\,\,\,}@{\,\,\,}c@{\,\,\,}}
				\toprule
				Symbol & Description & Value  \\
				\hline
				$m$ & \small{Equivalent mass of vehicle}  &       \small{2332 [kg]} \\
				$A_\text{f}$ & \small{Frontal projected area of vehicle}  &     \small{2.43 [$\mathrm{m^2}$]} \\
				$\rho_\text{a}$ & \small{Air density} &        \small{1.206  [$\mathrm{kg/m^3}$]} \\
				$C_\text{d}$ & \small{Air drag coefficient} &      \small{0.288 [-]} \\
				$C_\text{r}$ & \small{Rolling resistance coefficient} &     \small{0.0068 [-]} \\
				$\overline{F}_{\text{m}}$ & \small{Maximum traction force} & \small{10.1 [$\mathrm{kN}$]} \\
				$\overline{F}_{\text{b}}$ & \small{Maximum braking force}  & \small{10.1 [$\mathrm{kN}$]} \\
				$E_{\text{cap}}$ & \small{Battery's energy capacity} & \small{77.4 [$\mathrm{kWh}$]} \\
				$\underline{v}$ & \small{Minimum vehicle velocity} & \small{20 [$\mathrm{km/h}$]} \\
				$\bar{\zeta}$ & \small{Maximum SoC} & \small{100 [\%]} \\
				$\underline{\zeta}$ & \small{Minimum SoC} & \small{10 [\%]} \\
				$ v_{\text{init}} $ & \small{Initial velocity} & \small{30 [$\mathrm{km/h}$]} \\
				$\zeta_{\text{init}} $ & \small{Initial SoC} & \small{25 [\%]} \\
				$\zeta_{\text{final}} $ & \small{Final SoC} & \small{75 [\%]} \\
				$\overline{\tau}_\mathrm{ch} $ & \small{Maximum time spent at the station} & \small{60 [min]} \\
				$\underline{\tau}_\mathrm{ch} $ & \small{Waiting time for charging} & \small{5 [min]} \\
				\bottomrule
			\end{tabular}
		}
	\end{center}
\end{table}

\begin{figure}[t]
	\begin{center}
		\includegraphics[width=.4\textwidth]{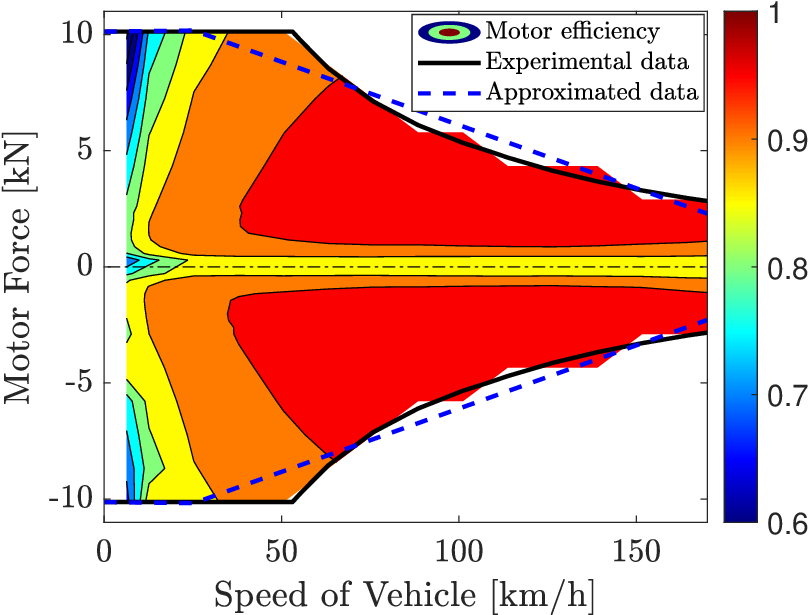}\vspace{-3mm}
	\end{center}
	\caption{Motor efficiency and force limits obtained from real-world experiments and approximated by numerical regressors.}
	\label{fig:motoreff_ftmax}
\end{figure}
\begin{figure}[t!]
	\centering
	\includegraphics[width=.37\textwidth]{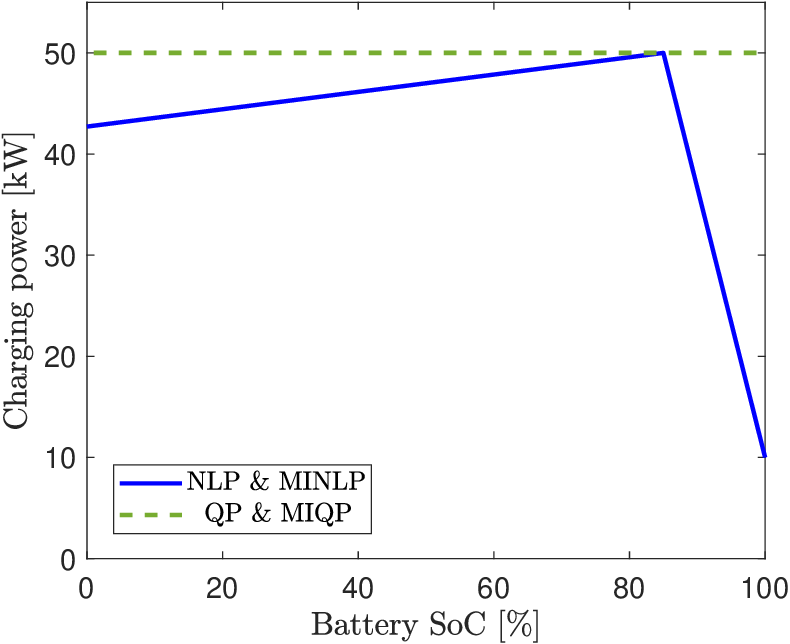}\vspace{-1mm}
	\caption{State-of-Charge (SoC)-dependent charging power profiles utilized in optimization-based speed and charging planning strategies. In (MI)QP formulations, a piecewise constant charging power is assumed, while (MI)NLP formulations incorporate charging power profiles that are piecewise affine with respect to the battery SoC.}
	\label{fig:chargingpower}
\end{figure}

\begin{table*}[h]
	\renewcommand{\arraystretch}{1.6}
	\captionsetup{justification=centering}
	\caption{Comparison of solution methods for simultaneous speed and charging planning without (NLP and QP) or with (MINLP, MIQP, and MIQP-greedy)  charging-stop constraints.}\vspace{-2mm}
	\begin{center}
		\centering
		{\Large
			\resizebox{1\textwidth}{!}{
					\begin{tabular}{|cc|ccccccc|ccccccc|ccccccc|}
						\hline
						\multicolumn{2}{|c|}{\textbf{Trip type}} & \multicolumn{7}{c|}{Case I : Short-range trip}  & \multicolumn{7}{c|}{Case II : Mid-range trip} & \multicolumn{7}{c|}{Case III : Long-range trip} \\ \hline
						\multicolumn{2}{|c|}{\multirow{3}{*}{\textbf{Method}}}                                                                 & \multicolumn{2}{c|}{\begin{tabular}[c]{@{}c@{}}Non-charging stop\\ constraints\end{tabular}} & \multicolumn{5}{c|}{\begin{tabular}[c]{@{}c@{}}With charging stop constraints\end{tabular}} & \multicolumn{2}{c|}{\begin{tabular}[c]{@{}c@{}}Non-charging stop\\ constraints\end{tabular}} & \multicolumn{5}{c|}{\begin{tabular}[c]{@{}c@{}}With charging stop constraints\end{tabular}} & \multicolumn{2}{c|}{\begin{tabular}[c]{@{}c@{}}Non-charging stop\\ constraints\end{tabular}} & \multicolumn{5}{c|}{\begin{tabular}[c]{@{}c@{}}With charging stop constraints\end{tabular}} \\ \cline{3-23}
						\multicolumn{2}{|c|}{}                                                                                                          & \multicolumn{1}{c|}{NLP}                        & \multicolumn{1}{c|}{QP}                        & \multicolumn{1}{c|}{MINLP}     & \multicolumn{2}{c|}{MIQP}                                                                                                                                               & \multicolumn{2}{c|}{\begin{tabular}[c]{@{}c@{}}MIQP-greedy\\ (solving QP subproblemss)\end{tabular}} & \multicolumn{1}{c|}{NLP}                        & \multicolumn{1}{c|}{QP}                        & \multicolumn{1}{c|}{MINLP}     & \multicolumn{2}{c|}{MIQP}                                                                                                                                               & \multicolumn{2}{c|}{\begin{tabular}[c]{@{}c@{}}MIQP-greedy\\ (solving QP subproblemss)\end{tabular}} & \multicolumn{1}{c|}{NLP}                        & \multicolumn{1}{c|}{QP}                        & \multicolumn{1}{c|}{MINLP}     & \multicolumn{2}{c|}{MIQP}                                                                                                                                               & \multicolumn{2}{c|}{\begin{tabular}[c]{@{}c@{}}MIQP-greedy\\ (solving QP subproblemss)\end{tabular}}              \\ \hline
						\multicolumn{2}{|c|}{\textbf{\begin{tabular}[c]{@{}c@{}}Number of charges\\$N_\mathrm{ch}$\end{tabular}}}                                                                                                                 & \multicolumn{2}{c|}{\begin{tabular}[c]{@{}c@{}}\LARGE{5}\\[-0.5em](all stations)\end{tabular}}                                                                           & \multicolumn{5}{c|}{\LARGE{2}}                                                                                                                                                                                                                                                                                          & \multicolumn{2}{c|}{\begin{tabular}[c]{@{}c@{}}\LARGE{12}\\[-0.5em](all stations)\end{tabular}}                                                                          & \multicolumn{5}{c|}{\LARGE{2}}                                                                                                                                                                                                                                                         & \multicolumn{2}{c|}{\begin{tabular}[c]{@{}c@{}}\LARGE{19}\\[-0.5em](all stations)\end{tabular}}                                                                          & \multicolumn{5}{c|}{\LARGE{3}}                                                                                                                                                                                                                                                        \\ \hline
						\multicolumn{2}{|c|}{\multirow{3}{*}{\textbf{Solver}}}                                                                          & \multicolumn{2}{c|}{open-source}                                                                                                  & \multicolumn{1}{c|}{open-source}                                                   &
						\multicolumn{1}{c|}{open-source}                                                   & \multicolumn{1}{c|}{commercial}                                                    & \multicolumn{1}{c|}{open-source}                             & commercial                            & \multicolumn{2}{c|}{open-source}                                                                                                  & \multicolumn{1}{c|}{open-source}                                                   &
						\multicolumn{1}{c|}{open-source}                                                   & \multicolumn{1}{c|}{commercial}                                                    & \multicolumn{1}{c|}{open-source} & commercial  & \multicolumn{2}{c|}{open-source}                                                                                                  & \multicolumn{1}{c|}{open-source}                                                   &
						\multicolumn{1}{c|}{open-source}                                                   & \multicolumn{1}{c|}{commercial}                                                    & \multicolumn{1}{c|}{open-source} & commercial \\ \cline{3-23} 
						\multicolumn{2}{|c|}{}                                                                                                          & \multicolumn{1}{c|}{IPOPT}                      & \multicolumn{1}{c|}{ECOS}                      & \multicolumn{1}{c|}{BONMIN}    & \multicolumn{1}{c|}{\begin{tabular}[c]{@{}c@{}}BONMIN\\ SCIP\end{tabular}}         & \multicolumn{1}{c|}{\begin{tabular}[c]{@{}c@{}}GUROBI\\ CPLEX\end{tabular}}        & \multicolumn{1}{c|}{ECOS}                                    & GUROBI                                & \multicolumn{1}{c|}{IPOPT}                      & \multicolumn{1}{c|}{ECOS}                      & \multicolumn{1}{c|}{BONMIN}    & \multicolumn{1}{c|}{\begin{tabular}[c]{@{}c@{}}BONMIN\\ SCIP\end{tabular}}         & \multicolumn{1}{c|}{\begin{tabular}[c]{@{}c@{}}GUROBI\\ CPLEX\end{tabular}}        & \multicolumn{1}{c|}{ECOS}        & GUROBI      & \multicolumn{1}{c|}{IPOPT}                      & \multicolumn{1}{c|}{ECOS}                      & \multicolumn{1}{c|}{BONMIN}    & \multicolumn{1}{c|}{\begin{tabular}[c]{@{}c@{}}BONMIN\\ SCIP\end{tabular}}         & \multicolumn{1}{c|}{\begin{tabular}[c]{@{}c@{}}GUROBI\\ CPLEX\end{tabular}}        & \multicolumn{1}{c|}{ECOS}        & GUROBI     \\ \hhline{|=======================|}
						\multicolumn{1}{|c|}{\multirow{7.5}{*}{\rotatebox{90}{\makebox[1.9cm][c]{\textbf{Solution}}}}} & \textbf{\begin{tabular}[c]{@{}c@{}}Trip-time\\$t_N$\\ {[}min{]}\end{tabular}}     & \multicolumn{1}{c|}{243.12}                     & \multicolumn{1}{c|}{233.51}                    & \multicolumn{1}{c|}{224.45}                 & \multicolumn{1}{c|}{\begin{tabular}[c]{@{}c@{}}218.52\\ 218.47\end{tabular}}       & \multicolumn{1}{c|}{\begin{tabular}[c]{@{}c@{}}218.52\\ 218.52\end{tabular}}       & \multicolumn{1}{c|}{218.51}                                  & 218.52                                & \multicolumn{1}{c|}{423.16}                     & \multicolumn{1}{c|}{419.29}                    & \multicolumn{1}{c|}{377.36}                 & \multicolumn{1}{c|}{\begin{tabular}[c]{@{}c@{}}369.30\\ 368.87\end{tabular}}       & \multicolumn{1}{c|}{\begin{tabular}[c]{@{}c@{}}369.30\\ 369.30\end{tabular}}       & \multicolumn{1}{c|}{369.23}      & 369.30      & \multicolumn{1}{c|}{677.97}                     & \multicolumn{1}{c|}{675.92}                    & \multicolumn{1}{c|}{601.50}                 & \multicolumn{1}{c|}{\begin{tabular}[c]{@{}c@{}}595.93\\ 595.53\end{tabular}}       & \multicolumn{1}{c|}{\begin{tabular}[c]{@{}c@{}}595.93\\ 595.93\end{tabular}}       & \multicolumn{1}{c|}{595.92}      & 595.93     \\ \cline{2-23} 
						\multicolumn{1}{|c|}{}                                   & \textbf{\begin{tabular}[c]{@{}c@{}}\ Net charging time\ \\$\tau_\mathrm{ch}-N_\mathrm{ch}\underline{\tau}_\mathrm{ch}$\\ {[}min{]}\end{tabular}} & \multicolumn{1}{c|}{78.83}                      & \multicolumn{1}{c|}{74.10}                     & \multicolumn{1}{c|}{78.68}                  & \multicolumn{1}{c|}{\begin{tabular}[c]{@{}c@{}}74.11\\ 74.18\end{tabular}}         & \multicolumn{1}{c|}{\begin{tabular}[c]{@{}c@{}}74.11\\ 74.11\end{tabular}}         & \multicolumn{1}{c|}{74.09}                                   & 74.09                                 & \multicolumn{1}{c|}{102.77}                     & \multicolumn{1}{c|}{101.59}                    & \multicolumn{1}{c|}{107.24}                 & \multicolumn{1}{c|}{\begin{tabular}[c]{@{}c@{}}101.60\\ 101.82\end{tabular}}       & \multicolumn{1}{c|}{\begin{tabular}[c]{@{}c@{}}101.60\\ 101.60\end{tabular}}       & \multicolumn{1}{c|}{101.65}      & 101.57      & \multicolumn{1}{c|}{136.19}                     & \multicolumn{1}{c|}{137.33}                    & \multicolumn{1}{c|}{140.74}                 & \multicolumn{1}{c|}{\begin{tabular}[c]{@{}c@{}}137.34\\ 137.38\end{tabular}}       & \multicolumn{1}{c|}{\begin{tabular}[c]{@{}c@{}}137.34\\ 137.34\end{tabular}}       & \multicolumn{1}{c|}{137.27}      & 137.28     \\ \cline{2-23} 
						\multicolumn{1}{|c|}{}                                   & \textbf{\begin{tabular}[c]{@{}c@{}}Energy-consumption\\$E_\mathrm{mech}$\\ {[}MJ{]}\end{tabular}}  & \multicolumn{1}{c|}{120.78}                     & \multicolumn{1}{c|}{124.61}                    & \multicolumn{1}{c|}{124.58}                 & \multicolumn{1}{c|}{\begin{tabular}[c]{@{}c@{}}124.63\\ 124.92\end{tabular}}       & \multicolumn{1}{c|}{\begin{tabular}[c]{@{}c@{}}124.63\\ 124.63\end{tabular}}       & \multicolumn{1}{c|}{124.61}                                  & 124.63                                & \multicolumn{1}{c|}{229.31}                     & \multicolumn{1}{c|}{230.62}                    & \multicolumn{1}{c|}{229.41}                 & \multicolumn{1}{c|}{\begin{tabular}[c]{@{}c@{}}230.66\\ 231.53\end{tabular}}       & \multicolumn{1}{c|}{\begin{tabular}[c]{@{}c@{}}230.66\\ 230.66\end{tabular}}       & \multicolumn{1}{c|}{231.01}      & 230.66      & \multicolumn{1}{c|}{365.84}                     & \multicolumn{1}{c|}{368.47}                    & \multicolumn{1}{c|}{365.39}                 & \multicolumn{1}{c|}{\begin{tabular}[c]{@{}c@{}}368.51\\ 368.67\end{tabular}}       & \multicolumn{1}{c|}{\begin{tabular}[c]{@{}c@{}}368.51\\ 368.51\end{tabular}}       & \multicolumn{1}{c|}{368.47}      & 368.51     \\ \hline
						\multicolumn{2}{|c|}{\textbf{\begin{tabular}[c]{@{}c@{}}Sub-optimality\\ for $t_N$\end{tabular}}} & \multicolumn{1}{c|}{100.00 \%}                  & \multicolumn{1}{c|}{103.95 \%}                 & \multicolumn{1}{c|}{107.68 \%}              & \multicolumn{1}{c|}{\begin{tabular}[c]{@{}c@{}}110.12 \%\\ 110.14 \%\end{tabular}} & \multicolumn{1}{c|}{\begin{tabular}[c]{@{}c@{}}110.12 \%\\ 110.12 \%\end{tabular}} & \multicolumn{1}{c|}{110.12 \%}                               & 110.12 \%                             & \multicolumn{1}{c|}{100.00 \%}                  & \multicolumn{1}{c|}{100.91 \%}                 & \multicolumn{1}{c|}{110.82 \%}              & \multicolumn{1}{c|}{\begin{tabular}[c]{@{}c@{}}112.73 \%\\ 122.83 \%\end{tabular}} & \multicolumn{1}{c|}{\begin{tabular}[c]{@{}c@{}}112.73 \%\\ 112.73 \%\end{tabular}} & \multicolumn{1}{c|}{112.74 \%}   & 112.73 \%   & \multicolumn{1}{c|}{100.00 \%}                  & \multicolumn{1}{c|}{100.30 \%}                 & \multicolumn{1}{c|}{111.28 \%}              & \multicolumn{1}{c|}{\begin{tabular}[c]{@{}c@{}}112.10 \%\\ 112.16 \%\end{tabular}} & \multicolumn{1}{c|}{\begin{tabular}[c]{@{}c@{}}112.10 \%\\ 112.10 \%\end{tabular}} & \multicolumn{1}{c|}{112.10 \%}   & 112.10 \%  \\ \hline
						\multicolumn{2}{|c|}{\textbf{\begin{tabular}[c]{@{}c@{}}Computation time \\{[}sec{]}\end{tabular}}}                      & \multicolumn{1}{c|}{0.149}                      & \multicolumn{1}{c|}{0.031}                     & \multicolumn{1}{c|}{22.10}                 & \multicolumn{1}{c|}{\begin{tabular}[c]{@{}c@{}}21.41\\ 0.554\end{tabular}}         & \multicolumn{1}{c|}{\begin{tabular}[c]{@{}c@{}}0.043\\ 0.912\end{tabular}}         & \multicolumn{1}{c|}{1.325}                                   & 0.095                                 & \multicolumn{1}{c|}{0.456}                      & \multicolumn{1}{c|}{0.084}                     & \multicolumn{1}{c|}{108.34}                 & \multicolumn{1}{c|}{\begin{tabular}[c]{@{}c@{}}83.13\\ 1.694\end{tabular}}        & \multicolumn{1}{c|}{\begin{tabular}[c]{@{}c@{}}0.113\\ 0.443\end{tabular}}         & \multicolumn{1}{c|}{29.443}      & 1.081       & \multicolumn{1}{c|}{0.786}                      & \multicolumn{1}{c|}{0.191}                     & \multicolumn{1}{c|}{3717.56}                & \multicolumn{1}{c|}{\begin{tabular}[c]{@{}c@{}}1429.92\\ 6.179\end{tabular}}      & \multicolumn{1}{c|}{\begin{tabular}[c]{@{}c@{}}0.519\\ 1.630\end{tabular}}                          & \multicolumn{1}{c|}{525.356}     & 26.934     \\ \hline
						\multicolumn{2}{|c|}{\textbf{Objective value}}                                                                        & \multicolumn{1}{c|}{14673.9}                    & \multicolumn{1}{c|}{-819.625}                  & \multicolumn{1}{c|}{13543.4}                & \multicolumn{1}{c|}{\begin{tabular}[c]{@{}c@{}}-909.609\\ -948.273\end{tabular}}   & \multicolumn{1}{c|}{\begin{tabular}[c]{@{}c@{}}-909.599\\ -909.599\end{tabular}}   & \multicolumn{1}{c|}{-909.630}                                & -909.599                              & \multicolumn{1}{c|}{25547.0}                    & \multicolumn{1}{c|}{-2237.07}                  & \multicolumn{1}{c|}{22795.5}                & \multicolumn{1}{c|}{\begin{tabular}[c]{@{}c@{}}-2537.02\\ -2615.24\end{tabular}}   & \multicolumn{1}{c|}{\begin{tabular}[c]{@{}c@{}}-2537.01\\ -2537.01\end{tabular}}   & \multicolumn{1}{c|}{-2562.20}    & -2537.01    & \multicolumn{1}{c|}{40855.1}                    & \multicolumn{1}{c|}{-3803.03}                  & \multicolumn{1}{c|}{36263.0}                & \multicolumn{1}{c|}{\begin{tabular}[c]{@{}c@{}}-4282.98\\ -4368.94\end{tabular}}   & \multicolumn{1}{c|}{\begin{tabular}[c]{@{}c@{}}-4282.97\\ -4282.97\end{tabular}}   & \multicolumn{1}{c|}{-4283.14}    & -4282.97   \\ \hline
					\end{tabular}
				}
		}
	\end{center}
	\label{table:solution}
\end{table*}

To validate the feasibility of the proposed MIQP method for $S^{2}CP$ developed from convex-OCP, simulation environments were constructed. The vehicle model parameters were obtained from a Hyundai IONIQ 5 and are listed in Table~\ref{table:vehicle}. The initial and final SoC of the EV's battery were assumed as 25\% and 75\%, respectively.

In this study, the three different case studies, referred to as Cases I, II, and III, are categorized based on driving distance, with route information shown in Fig.~\ref{fig:tripinfo}. The driving scenarios considered are as follows: Incheon-Pyeongchang (short-distance), Incheon-Busan (mid-distance), and Incheon-Busan-Haenam (long-distance) trips in Korea. As discussed in Section~\ref{subsec:constraints}, the speed constraints, including maximum speed $\overline{v}$, minimum speed $\underline{v}$, and average speed $v_\mathrm{avg}$ with a margin $\sigma$ of 10 km/h, as well as road elevation and the resulting road slope $\alpha$, are incorporated into the model. Furthermore, the number and locations of charging stations are included in the analysis. Real-time V2X driving condition data were utilized, provided by publicly available APIs such as Google Maps~\cite{GoogleMapsAPI}, SK TMAP~\cite{TmapAPI}, and the Korea Public Data Portal~\cite{KoeraPublicAPI}.

We conducted simulations of open-loop MPC using the four optimal control problems presented in Section~\ref{subsec:method_mip}. In the solution methods for NLP~\eqref{eq:nlp_s2cp} and MINLP~\eqref{eq:minlp}, the charging power $P_{\mathrm{ch},k}(\zeta)$ is the assumed profile, as shown in Fig. ~\ref{fig:chargingpower} referring to~\cite{FORCESPRO}, with the motor efficiency 2D map $\eta(v,F_\mathrm{m})$ and traction force limit $\overline{F}^\mathrm{cp}_{\mathrm{m},k}(v_k)$ shown in Fig.~\ref{fig:motoreff_ftmax}. In contrast, in the cases of QP~\eqref{eq:qp_s2cp} and MIQP~\eqref{eq:miqp}, we assumed that the charging power $P_\mathrm{ch}$ and motor efficiency $\eta$ had constant values of 50 kW and 0.9, respectively, for convexity. Subsequently, the traction force limit $\overline{F}^\mathrm{cp}_{\mathrm{m},k}(x_k)$ was approximated as an affine function of $x_k:=v_k^2$, as illustrated in Fig.~\ref{fig:motoreff_ftmax}.

In particular, our implementation includes the MIQP-greedy approach, which involves solving all the QP subproblems \eqref{eq:qp_s2cp} of the MIQP while considering the number of charging stations and then searching for the solution with the minimum objective value. The number of charges $N_\mathrm{ch}$ obtained from \eqref{eq:minNch} is two for Cases I and II and three for Case III. Therefore, the numbers of QPs to be solved to obtain a solution for the MIQP-greedy approach are ${}_5\mathrm{C}_2=10$, ${}_{12}\mathrm{C}_2=66$, and ${}_{19}\mathrm{C}_3=969$.

For the simulation in this study, the open-source solvers IPOPT~\cite{IPOPT}, ECOS~\cite{ECOS}, and BONMIN~\cite{BONMIN} were used for the NLP, QP, and MINLP algorithms, respectively. For the MIQP problem in $S^2CP$, the open-source solvers BONMIN and SCIP~\cite{SCIP} were applied, and the high-performance commercial solvers GUROBI~\cite{GUROBI} and CPLEX~\cite{CPLEX} were used for verification. Additionally, to solve the QP subproblems in the MIQP-greedy method, ECOS and GUROBI were compared. The MPC simulations for each driving scenario were conducted in a MATLAB environment using either CasADi~\cite{CASADI} or custom solver interfaces, depending on the solver. All computations were performed on a laptop equipped with an Intel Ultra 9 CPU 2.3 GHz (up to 5.1 GHz) and 32 GB of RAM.

\begin{figure}[ht!]
	\begin{subfigure}{\linewidth}
		\centering
		\includegraphics[width=\textwidth]{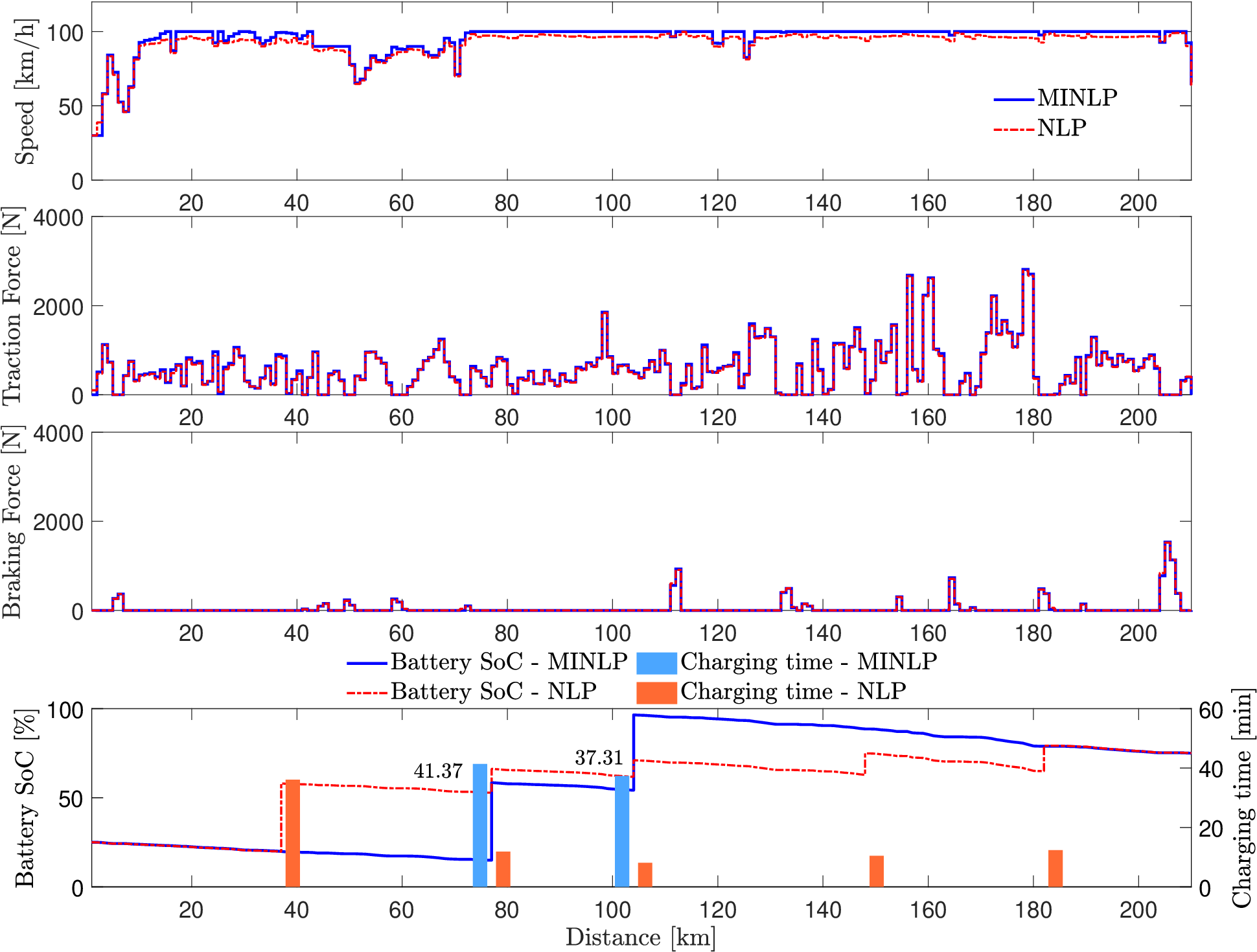}\vspace{-1mm}
		\caption{\small Case I: Trip for Incheon $\rightarrow$ Pyeongchang.}
		\label{fig:nlp_minlp_trip1}
	\end{subfigure}
	\vfill
	\vspace{4mm}
	\begin{subfigure}{\linewidth}
		\centering
		\includegraphics[width=\textwidth]{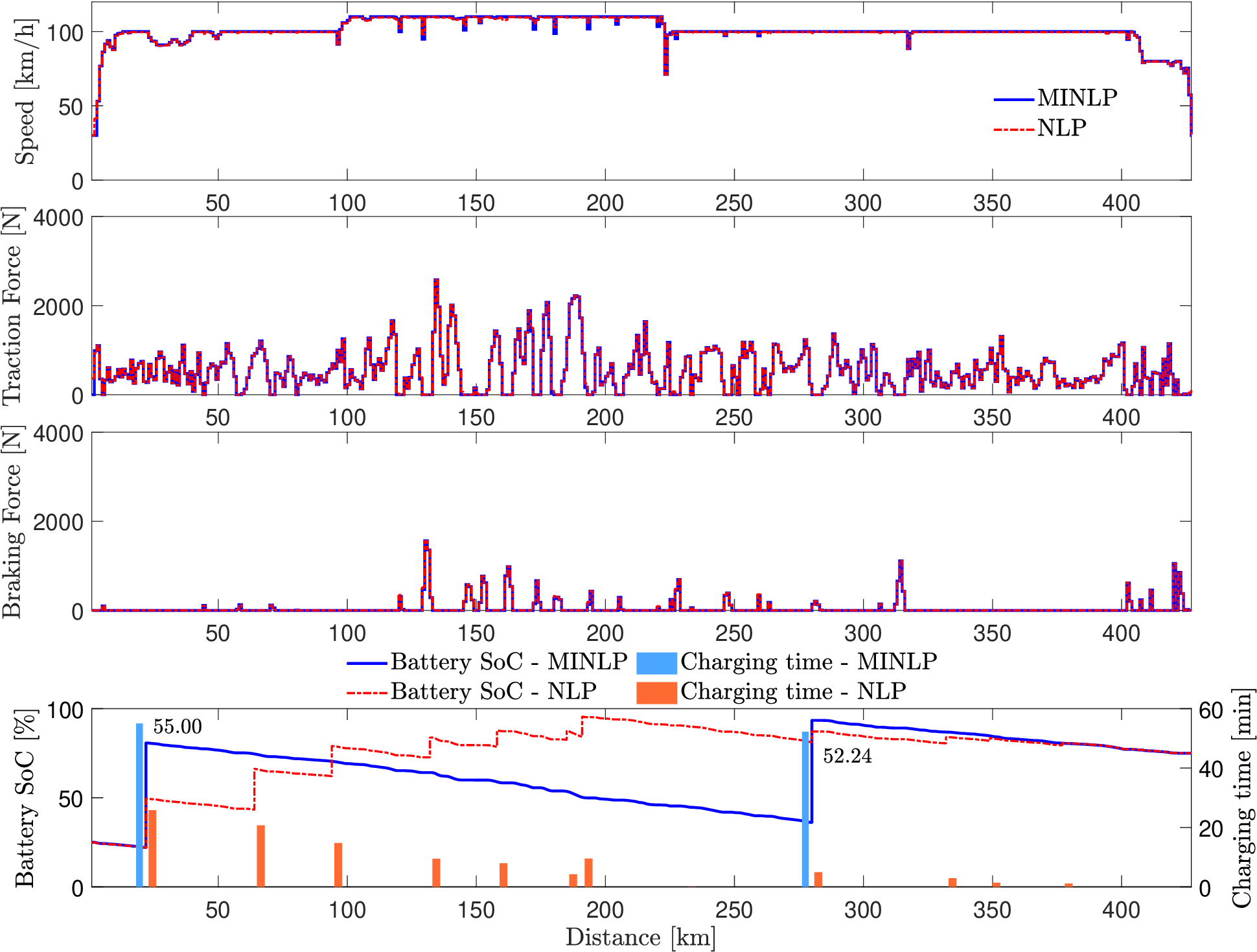}\vspace{-1mm}
		\caption{\small Case II: Trip for Incheon $\rightarrow$ Busan.}
		\label{fig:nlp_minlp_trip2}
	\end{subfigure}
	\vfill
	\vspace{4mm}
	\begin{subfigure}{\linewidth}
		\centering
		\includegraphics[width=\textwidth]{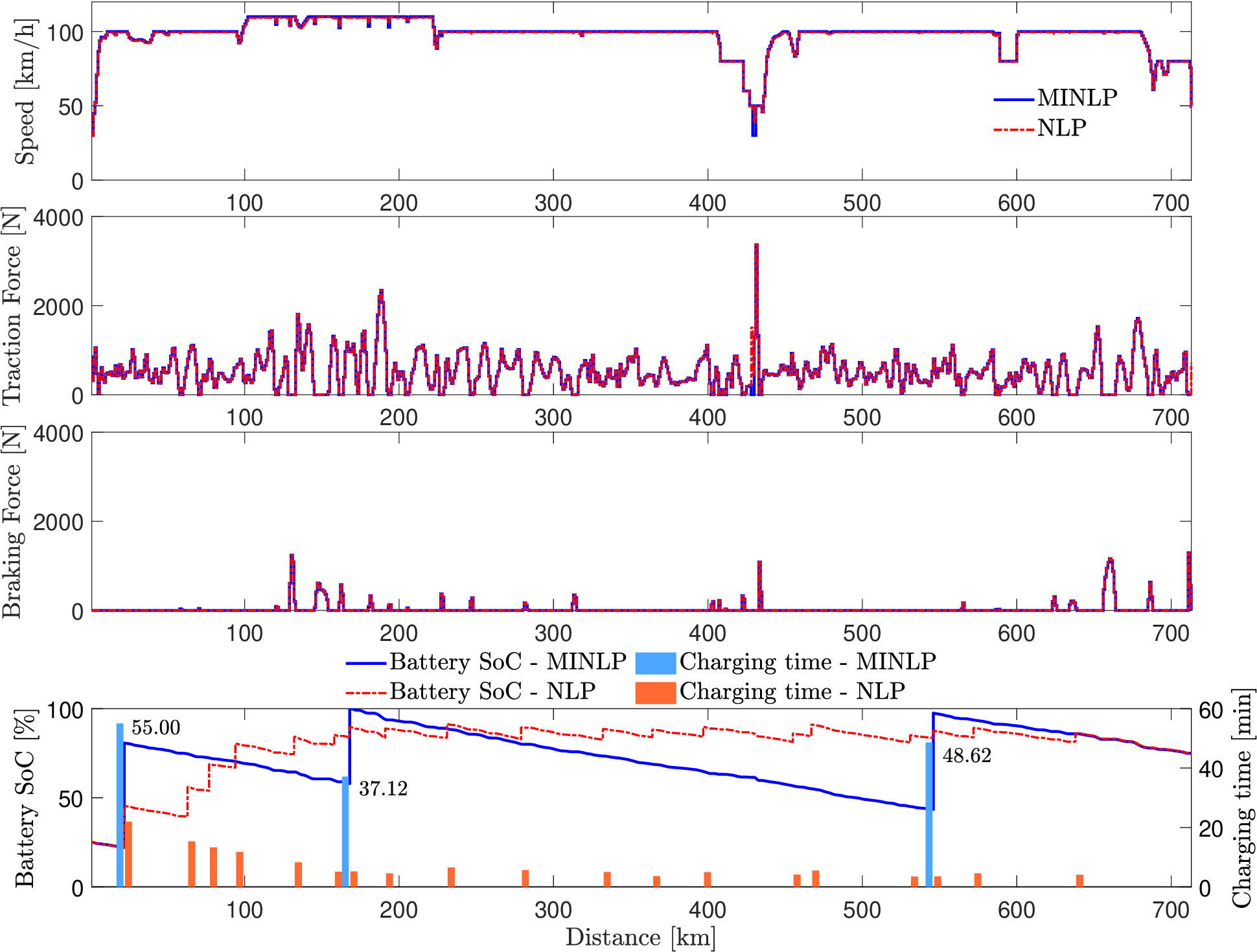}\vspace{-1mm}
		\caption{\small Case III: Trip for Incheon $\rightarrow$ Busan $\rightarrow$ Haenam.}
		\label{fig:nlp_minlp_trip3}
	\end{subfigure}
	\caption{Solution trajectories for speed and charging strategies obtained from MPC using NLP and MINLP formulations.}
	\label{fig:nlp_minlp}
\end{figure}

\begin{figure}[ht!]
	\begin{subfigure}{\linewidth}
		\centering
		\includegraphics[width=\textwidth]{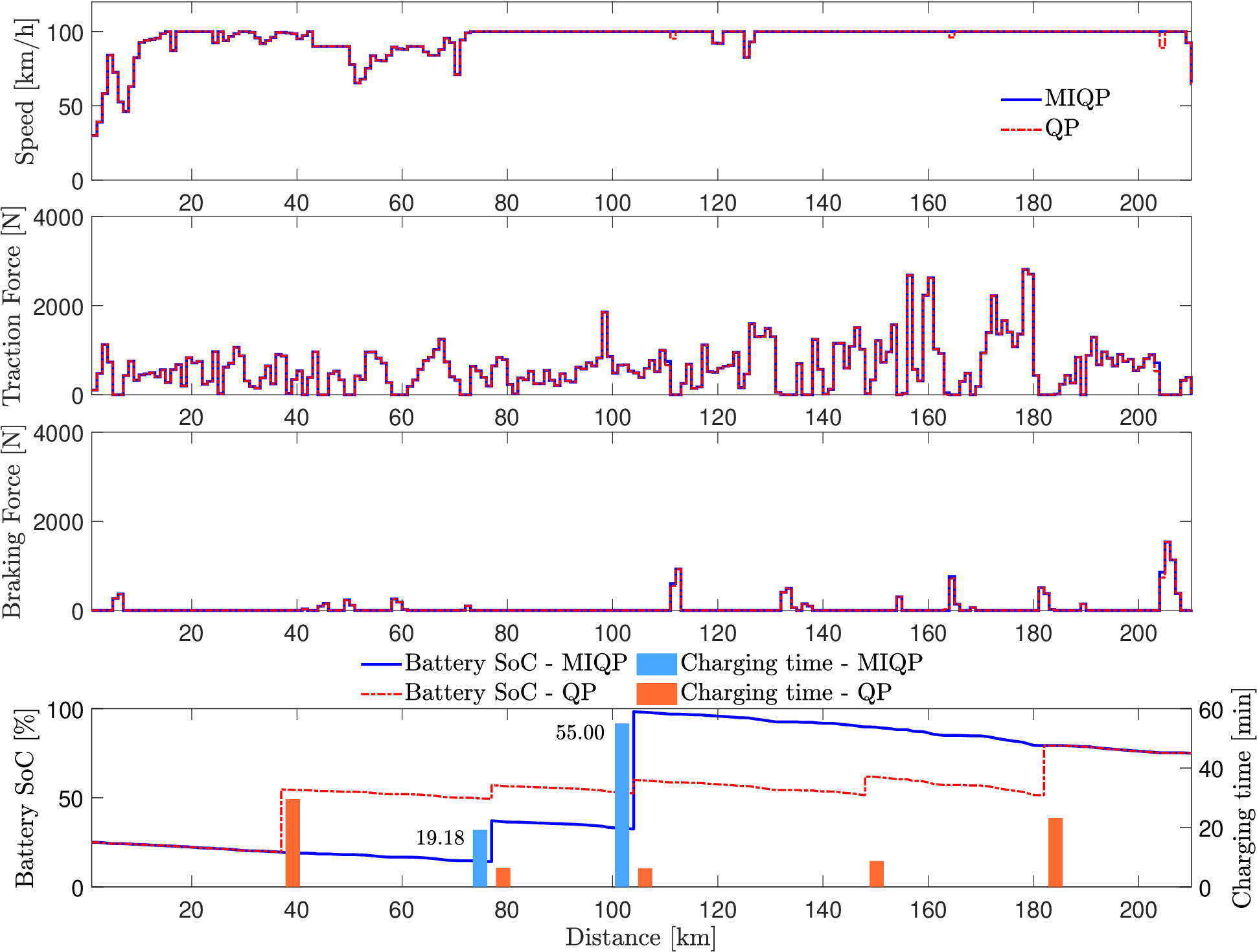}\vspace{-1mm}
		\caption{\small Case I: Trip for Incheon $\rightarrow$ Pyeongchang.}
		\label{fig:qp_miqp_trip1}
	\end{subfigure}
	\vfill
	\vspace{4mm}
	\begin{subfigure}{\linewidth}
		\centering
		\includegraphics[width=\textwidth]{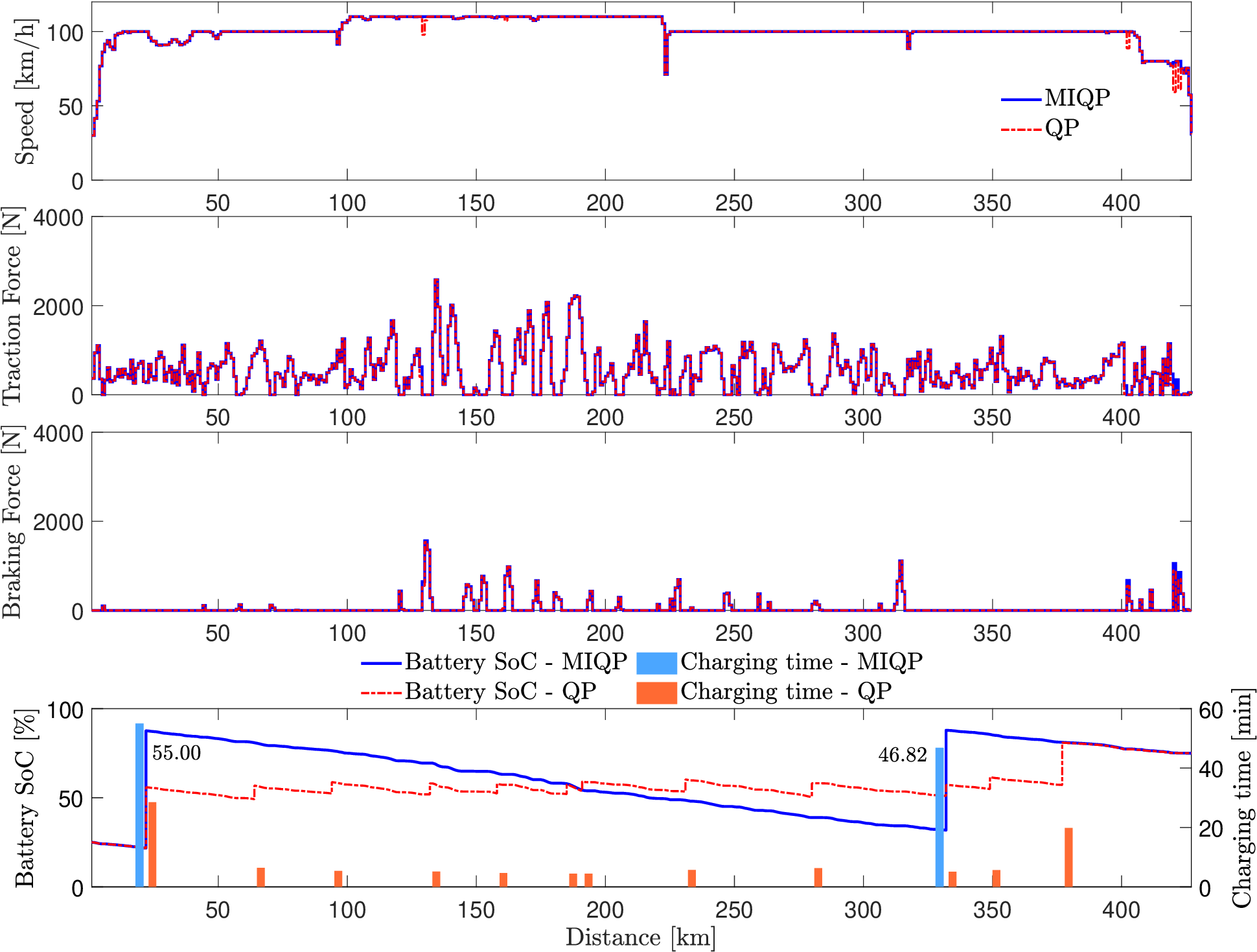}\vspace{-1mm}
		\caption{\small Case II: Trip for Incheon $\rightarrow$ Busan.}
		\label{fig:qp_miqp_trip2}
	\end{subfigure}
	\vfill
	\vspace{4mm}
	\begin{subfigure}{\linewidth}
		\centering
		\includegraphics[width=\textwidth]{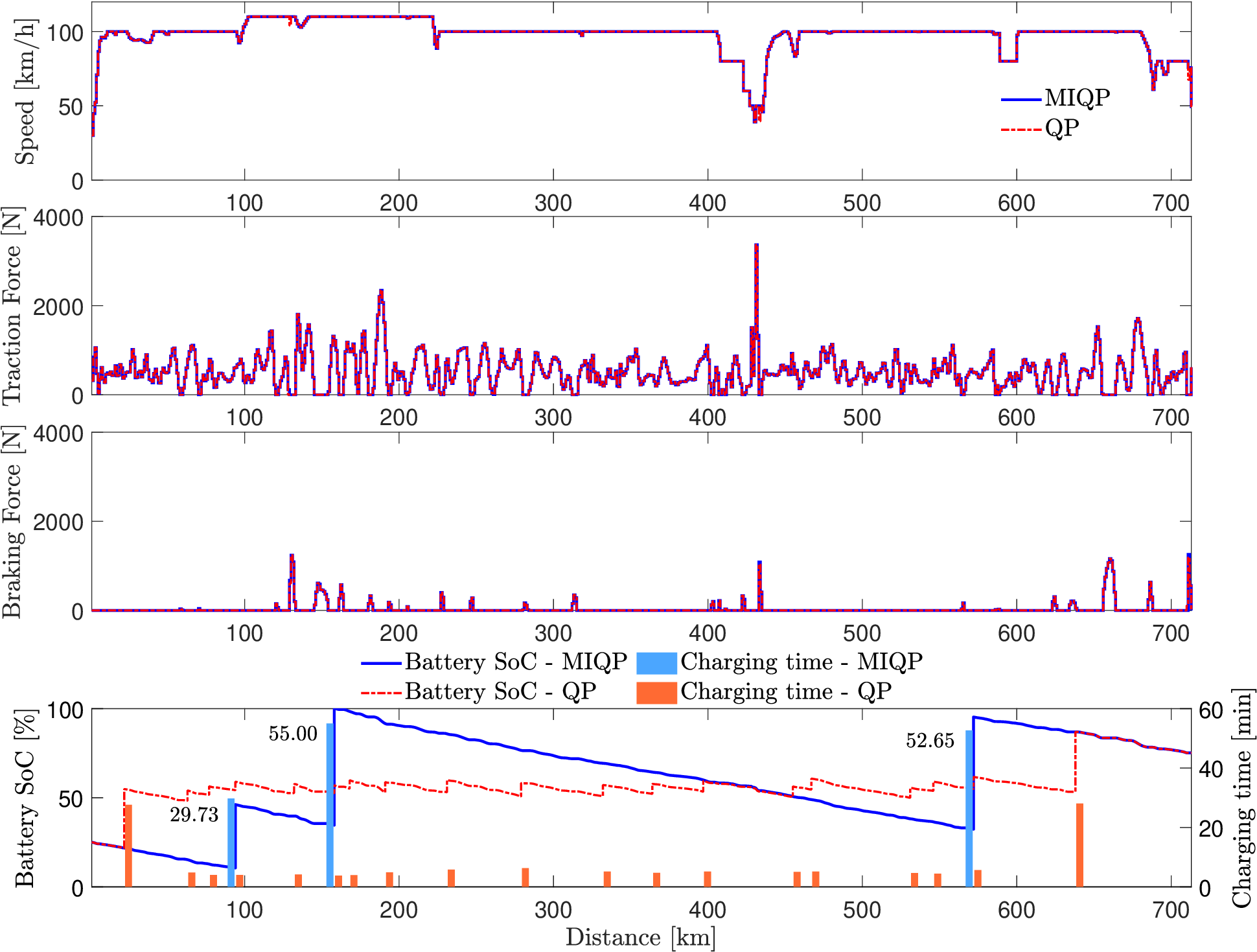}\vspace{-1mm}
		\caption{\small Case III: Trip for Incheon $\rightarrow$ Busan $\rightarrow$ Haenam.}
		\label{fig:qp_miqp_trip3}
	\end{subfigure}
	\caption{Solution trajectories for speed and charging strategies obtained from MPC using QP and MIQP formulations.}
	\label{fig:qp_miqp}
\end{figure}

\begin{figure}[t]
	\centering
	\includegraphics[width=.4\textwidth]{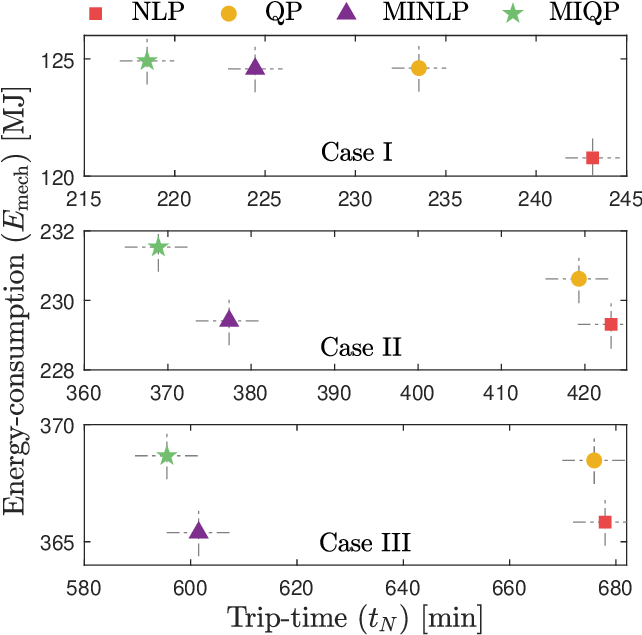}
	\caption{Comparison of multi-objective trade-off performances for four different formulations: (a) Planning without charging-stop constraints (NLP and QP) and (b) Planning with charging-stop constraints (MINLP and MIQP).}
	\label{fig:pareto}
	\vspace{4mm}
	\centering
	\includegraphics[width=.4\textwidth]{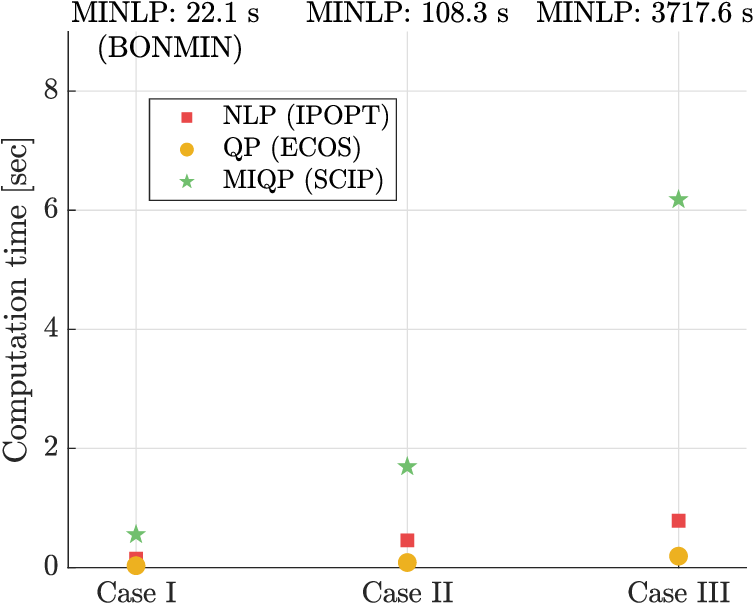}
	\caption{Comparison of computation times for NLP, QP, MINLP, and MIQP across three trip scenarios. Computation times for MINLP, solved using BONMIN, are not displayed in the figure but are indicated as numerical values at the top.}
	\label{fig:computationtime}
\end{figure}

\subsection{Case I: Short-distance Trip $(\leq 250\:\rm{km})$}
For the route from Incheon to Pyeongchang, the optimal speed plan and corresponding traction and braking force trajectories can be observed from the results in Figs.~\ref{fig:nlp_minlp_trip1} and ~\ref{fig:qp_miqp_trip1}. The obtained solution satisfied one of our objectives, with a final SoC of 75\%. Additionally, the battery SoC decreased owing to energy consumption, and the optimal charging plan to meet the constraints given in the OCP was also determined. The proposed speed plan was affected by speed conditions, as shown in Fig.~\ref{fig:tripinfo_trip1}, where the vehicle maintains its speed close to the maximum speed limit to minimize travel time. In addition, the traction force trajectory demonstrates that the optimal control input is computed to minimize energy consumption.

As discussed in Section~\ref{subsec:traclimit}, the physical constraint on the vehicle forces $F_\mathrm{m} \!\cdot\! F_\mathrm{b} = 0$ must be satisfied. Because the cost functions in our OCPs include both traction and braking forces, the simulation results naturally satisfy this condition.
In addition, because regenerative braking is not considered, the occurrence of a braking force in our OCPs indicates a loss in terms of energy or speed. However, in segments where the road slope increases sharply, a braking force is applied to satisfy the maximum road speed limit and the upper speed limit imposed by the surrounding traffic. Each method is compared with reference to the solution trajectories shown in Figs.~\ref{fig:nlp_minlp_trip1} and~\ref{fig:qp_miqp_trip1}.

\subsubsection{NLP vs. QP}
To assess the fidelity of the convexified QP problem derived from NLP, we compare the differences resulting from the linearization of model dynamics through simulation results. As shown in Table~\ref{table:solution}, the net charging time $\tau_\mathrm{ch}-N_\mathrm{ch}\underline{\tau}_\mathrm{ch}$ with respect to $N_\mathrm{ch}$ of 2 and total trip time are larger in the NLP than that in the QP. This is due to the fact that, as seen in Fig.~\ref{fig:chargingpower}, the charging power in the QP was fixed at a maximum value of 50 kW based on the NLP profile. Differences in the weighting factors of the cost function terms inherent to the problem formulations also contributed to this discrepancy.

The motor efficiency $\eta_k$ in the QP simulation was fixed at 90\%, as described previously. In contrast, $\eta(v,F_m)$ from the NLP solution's optimal speed profile exhibits a higher value of approximately 95\% throughout the trajectory of the trip. Consequently, the energy consumption $E_\mathrm{mech}$ was lower in the NLP than that in the QP. In terms of charging planning, both methods resulted in similar charging times for most charging stations, except for the final station.

\subsubsection{NLP vs. MINLP}
Notably, the charging plan that satisfies the battery SoC constraint shows a difference in the charging instances between NLP and MINLP. As seen in Fig.~\ref{fig:nlp_minlp_trip1}, the EV visits all five available charging stations using the NLP solution, whereas the plan obtained from the MINLP suggests visiting only the 2nd and 3rd stations with charging times of 41.37 min and 37.31 min, respectively.

The net total charging time for Case I from the solutions of MINLP and NLP are 78.68 min and 78.83 min, respectively, while the energy consumption $E_\mathrm{mech}$ is 124.58 MJ and 120.78 MJ, respectively. This shows that although the energy consumption of the MINLP solution is slightly higher, the charging time decreases by a corresponding amount owing to the opposite effect. However, because the number of charges is reduced, the effect of the minimum waiting time $\underline{\tau}_\mathrm{ch}$ at the stations introduced in Section~\ref{subsec:method_mip} results in a lower expected total trip time $t_N$. Specifically, the total trip time is 224.45 min for the MINLP solution, which is shorter than the 243.12 min for the NLP solution.

\subsubsection{QP vs. MIQP}
The results from QP and the MIQP developed from it were similar to the relationship between NLP and MINLP described earlier. In terms of charge planning, the constraint for the number of charges was the same as in the previous two methods, with five and two stations, respectively. Speed planning yielded approximately identical profiles; accordingly, the required force profiles were not significantly different. As shown in the MIQP solution in Table~\ref{table:solution}, a slight difference exists between the results from each solver. The trajectory of the MIQP shown in Fig.~\ref{fig:qp_miqp} shows only the SCIP, which is an open-source solver specialized for MIQP. Among the various characteristics of the MIQP solution, a maximum possible charging time of 55 min was derived at the 3rd station with respect to $\overline{\tau}_\mathrm{ch}-\underline{\tau}_\mathrm{ch}$.

In the separately considered greedy algorithm of the MIQP, an infeasible situation occurred in one of the ten QP subproblems, where charging was considered only at the 4th and 5th charging stations. This is because the battery SoC of the electric vehicle dropped below the minimum constraint of 10\% before reaching the 4th station. At this point, it becomes infeasible to find a solution that minimizes the battery energy consumption while satisfying the lower speed bound, as indicated in~\eqref{eq:vavg_const}, $v_\mathrm{avg}-\sigma \leq v$. In all other QP subproblemss, the feasible solutions from the GUROBI and ECOS solvers were similar to the other MIQP solutions.

\subsection{Case II: Mid-distance Trip $(250\sim500\:\rm{km})$}
In the Incheon-Busan trip, profiles of the optimal speed and charging are shown in Figs.~\ref{fig:nlp_minlp_trip2} and~\ref{fig:qp_miqp_trip2}. When comparing the relationships between the MPC solutions for speed and traction and braking forces across the four algorithms, the results were similar to those for Case I. In addition, regarding the plan for the battery SoC and charging time, the results from the NLP and QP algorithms showed that all charging stations were visited, as in Case I. However, more stations were found to complete charging in less time.
Alternatively, the results from MINLP and MIQP satisfy the constraint for $N_\mathrm{ch}$ of 2, highlighting the characteristic of MIP that allows for more efficient charging planning. The specific values are listed in Table~\ref{table:solution}, where it can be seen that the total trip-time difference between NLP and MINLP is 45.80 min, and between QP and MIQP (SCIP solver) is 50.42 min, both showing a larger gap compared to the previously simulated results from Case I.

The MIQP-greedy MPC calculated by both the GUROBI and ECOS solvers excluded 45 infeasible QP subproblems from the candidates for global solutions, leading to the derived solution. Additionally, the results were approximately identical to those obtained from the MIQP solver using the branch-and-bound exploration method.

\subsection{Case III: Long-distance Trip $(\geq 500\:\rm{km})$}
Similar outcomes between the algorithms were observed in the simulation of the long-distance driving routes from Incheon to Busan to Haenam, as shown in Figs.~\ref{fig:nlp_minlp_trip3} and~\ref{fig:qp_miqp_trip3}. Both the range of average speed and speed limit conditions on the road were satisfied, showing consistent behavior in the speed plan aimed at minimizing the energy consumption and trip time. In particular, because Case III was an extended simulation of the same driving route as in Case II, it closely matched the results up to 427 km, as shown in Figs.~\ref{fig:nlp_minlp_trip2} and~\ref{fig:qp_miqp_trip2}.
The number of charging instances in both the NLP and QP methods was 19 for all stations, whereas the MIP methods yielded results satisfying the minimum charging count $N_\mathrm{ch}$ of three. Consequently, the trip-time difference between NLP and MINLP was 76.47 min, and between QP and MIQP was 80.39 min, both showing a greater difference compared to that in Case II.

For the MIQP-greedy method, after filtering 657 infeasible combinations out of 969 QP subproblems, the obtained solution exhibited the same pattern as in the previous cases, showing results closely resembling those of MIQP-specific solvers.

\subsection{Optimality for Trip-time and Energy-consumption}\label{subsec:paretooptimality}
In this study, the primary metrics considered as cost functions are the trip time and energy consumption, as shown in equations \eqref{eq:tN}, \eqref{eq:J1}, and \eqref{eq:J2hat2}. The Pareto frontiers shown in Fig.~\ref{fig:pareto} illustrate the trade-off between $t_N$ and $E_\mathrm{mech}$. However, when categorizing the problems as MIP and non-MIP problems, the QP solution in Case I does not offer any improvement over the MINLP solution for both cost terms. Additionally, in both Cases II and III, the QP and NLP methods failed to provide tradeoff insights relative to the MINLP solution.

We now consider the battery energy consumption of the EV. First, all four solution methods in each case started with a 25\% SoC and precisely achieved a final SoC of 75\%. Second, the corresponding charging amount $\Delta\zeta_\mathrm{ch,req}$, similar to~\eqref{eq:maxenergycons}, is nearly consistent across all the solution methods within the same trip, as derived from the following equation:
\begin{equation}
	\Delta \zeta_\mathrm{ch,req} = \zeta_\mathrm{final} - \zeta_\mathrm{init} + \Delta \zeta_\mathrm{cons}\ .
\end{equation}
Therefore, our focus narrows to trip time, which is the aspect a driver perceives when following the plan proposed by $S^2CP$. Consequently, from the perspective of optimality, we compared the total trip-time solutions of QP, MINLP, and MIQP against NLP without charging-stop constraints, as described in Section~\ref{subsec:method_mip}.

As shown in Table~\ref{table:solution}, the optimality of $t_N$ for each case was evaluated based on the solutions obtained from NLP. Across all cases, the order of optimality relative to NLP followed the sequence designed for $S^2CP$: QP, MINLP, and MIQP with progressively higher optimality values. Specifically, the MIQP solution achieved optimality values of 110.14\%, 122.83\%, and 112.16\% for the three-trip cases, respectively, demonstrating the highest performance among the methods.

\subsection{Computation Time for Solvers}
In this study, system $S^2CP$ was implemented using open-source solvers to ensure that it can be widely adopted without being restricted by commercial licensing constraints. Therefore, we aimed to compare the solution times of the open-source solvers. In Fig.~\ref{fig:computationtime}, which shows the computational time relationships listed in Table~\ref{table:solution}, the solution time of the QP convexified from the NLP decreases by 79.2\%, 81.6\%, and 75.7\% in Cases I, II, and III, respectively. However, the addition of binary constraints in MINLP and MIQP led to an overall increase in computation time. In particular, when solving the MINLP problem using BONMIN, which is one of the most reliable nonconvex MIP open-source solvers, the computation time increased significantly compared with other methods. This is due to the solver's extensive exploration of the solution set of the nonlinear problem using a branch-and-bound approach. It is also highly sensitive to the initial estimate, which contributes to an increase in time. Consequently, when this solver is applied to the MIQP, the convergence speed is slower than that of other convex mixed-integer solvers.

By using SCIP instead of BONMIN, solutions were obtained in approximately 0.55, 1.69, and 6.18 s for each driving trip distance. Consequently, these convergence speeds were up to approximately 39, 49, and 231 times faster than BONMIN for Cases I, II, and III, respectively, significantly improving the real-time performance. For most MIQP problems considered, the commercial solvers derived accurate solutions in significantly less time than those of the open-source solvers. Notably, the state-of-the-art solver GUROBI demonstrated extremely fast computation times, solving all cases in less than 0.6 s. However, an important observation is that in terms of the solution quality, SCIP and BONMIN yielded approximately identical optimality values to those of the commercial solvers GUROBI and CPLEX. The optimality of the MIQP-greedy method, considered as global solutions, also indicates that these results are meaningful as an optimal plan for $S^2CP$ in an EV.

Additionally, when solving the QP and MIQP-greedy problems, we attempted to use the well-known open-source OSQP solver~\cite{OSQP}, but encountered issues where the inequality constraints were violated or state updates in the spatial domain steps did not properly reflect the given equalities of the model dynamics. This issue is likely due to the sensitivity to the size of the constraints inherent in the ADMM algorithm, which is used in OSQP, based on a first-order method. While the solver is an excellent tool for small-scale real-time MPCs with QP in embedded environments, our problem involves solving large-scale open-loop MPC problems with a relatively large number of constraints, which contribute to a lower solution accuracy.

In contrast, ECOS, which leverages the primal-dual interior point method (IPM) to directly solve the KKT system via the Newton iterations, demonstrated high accuracy. It was used to solve QP problems involving non-charging-stop constraints. Additionally, the open-source solver ECOS was applied as a benchmark against the commercial QP solver GUROBI to address MIQP-greedy problems. As a result, excluding the difference in computation time between the two types of solvers, similar global solutions were obtained from the minimum objective values of the QP subproblems.

\section{Discussion}\label{sec:discussion}

\subsection{Computational Complexity}
 The computational complexity of solving an MIQP using branch-and-bound algorithms is generally exponential in the worst-case scenarios. However, the practical performance often deviates significantly from this worst-case scenario owing to advancements in solver efficiency, tight bounding mechanisms, effective pruning strategies, and the application of heuristics. The actual computational effort required depends heavily on the specific problem instance and implementation of the branch-and-bound algorithm. In practice, real-world MIQP instances are frequently solved more efficiently than in theoretical worst-case analyses.

The key practical considerations for improving the computational efficiency include the following:  
\begin{itemize}  
\item
\emph{Embedded QP solvers:} Efficient QP solvers are critical since each node in the branch-and-bound tree involves solving a QP problem. High-performance solvers significantly reduce the computation time.  
\item
\emph{Heuristics:} Effective heuristics for generating initial feasible solutions and guiding the search process can significantly enhance solver performance by reducing the number of nodes explored.  
\item
\emph{Relaxations:} Linear relaxations of the MIQP are often used to compute lower bounds, facilitating pruning and improving convergence rates.  
\item
\emph{Computing time improvement using GPU-accelerated parallel computing:} Leveraging parallel computing to explore multiple branches simultaneously can reduce the overall computation time, particularly for large-scale problems.  
\end{itemize}  
By incorporating these techniques, the practical complexity of solving MIQP problems can be mitigated substantially, enabling the efficient resolution of instances relevant to real-world applications.  

\subsection{Nonlinearity of Battery Charging Curve}
\label{subsec:nonlinearcurve}
In this simulation, when the EV is connected to a charging station, we assume a fixed charging power, given by \({P_\mathrm{ch} \over E_\mathrm{cap}}\tau_\mathrm{ch}\), as described in~\eqref{eq:zeta_euler2}. However, in real-world scenarios, the charging power typically varies dynamically based on the battery's SoC and charging profile, which depend on the specific characteristics of the charger and battery system. Notably, the profile shown in Fig.~\ref{fig:chargingpower} was considered exclusively within the spatial domain. When an EV stops at a charging station, the charging energy demand can be more accurately modeled using a nonlinear approach that accounts for the variation in charging power as a function of the battery SoC and temperature level.

\subsection{Planning over Heterogeneous Spatial and Temporal Domains}
\label{subsec:spatialtemporal}
Two potential approaches can be considered to develop a high-accuracy EV driving strategy that incorporates the methods discussed in Section~\ref{subsec:nonlinearcurve}.  

The first approach involves implementing a sequential planner that solves the $S^2CP$ problem in the spatial domain using the MIQP method, followed by addressing the nonlinear time-domain problem associated with the charging profile at each station. This approach allows for an integrated solution by combining the spatial planning of routes and charging stops with the temporal optimization of charging durations.  

Alternatively, a more comprehensive and complex formulation can be devised as an MINLP problem. This method integrates the MIQP and NLP approaches to consider both the spatial and temporal domains simultaneously. Such a formulation allows for a unified optimization framework, potentially achieving greater accuracy and efficiency, albeit at the cost of increased computational complexity.  

Both approaches aim to balance practicality and precision and provide robust solutions for real-world EV planning scenarios.

\subsection{Optimality for Charging Planner}  
\label{subsec:optimality}  
As shown in Figs.~\ref{fig:nlp_minlp} and~\ref{fig:qp_miqp}, the speed-planning problem consistently yielded identical solutions across various cases. However, the charging strategies for EV trips exhibited distinct variations. Notably, even when the same MIQP method is used, different solvers may produce varying solutions for charging station selection. This discrepancy arises from the branch-and-bound process in MIQP, in which the differences in objective values between feasible solutions derived from QP subproblems are often negligible.  

As discussed in Section~\ref{subsec:nonlinearcurve}, the current assumptions are that all charging stations provide the same electrical power and that chargers are always available. To better align with real-world scenarios, incorporating constraints that account for the capacity of the maximum charging power at each station and the potential charger occupancy by other users enhances the optimality of the charging plan. V2X communication is a practical method for acquiring real-time information, thereby enabling more realistic and efficient charging strategies.

\section{Conclusions and Future Work}\label{sec:conclusion}
In this paper, we propose optimization-based strategies that integrate speed and charging planning for EVs, developed sequentially in the spatial domain. The proposed approaches successfully achieved a user-defined desired final energy state (SoC) across all trip scenarios—short, mid, and long-range driving—meeting the key objective of ADAS implementation. The original optimal control problem was formulated using NLP and subsequently simplified into a convex QP model by linearizing the model dynamics, thereby significantly reducing the computational complexity. The planning solution obtained through MPC produced comparable driving profiles while achieving a 4$\sim$5$\times$ reduction in computing time.  

To further enhance the framework, the model was extended to an MIQP problem, which optimally limited the total number of charging events and selected charging stations, along with their respective charging durations, to maximize user convenience. This led to the implementation of the $S^2CP$ planning system, which minimized the number of charging stops. Although slight differences in charging station indices were observed between the MINLP and MIQP solutions in the spatial domain (as discussed in Section~\ref{subsec:optimality}), the MIQP method demonstrated exceptional computational efficiency, completing long-range trip planning in approximately 6 s—231$\times$ faster than MINLP using open-source solvers—with negligible impact on the total trip time.  

As discussed in Section~\ref{subsec:spatialtemporal}, incorporating temporal planning into the MIQP framework provides opportunities for further improvement. Enhancing the model with realistic charging power profiles and considering the increase in battery temperature during charging could increase its accuracy and practicality. However, the nonconvex nature of the MINLP problem introduces substantial computational challenges, making simplification essential for real-time applications.  

Our future work will focus on redesigning the OCP in the temporal domain and prioritizing the optimization of the charging times at each station. To address the complexity of this problem, we plan to leverage the speed profiles and energy consumption predictions derived separately from the QP problem. Additionally, to dynamically update charging station availability during trips, we implemented an MPC approach with a reduced prediction horizon by dividing the total distance into manageable segments. This strategy aims to enhance the computational efficiency while preserving the quality and feasibility of solutions for real-world applications.


\balance

\bibliographystyle{IEEEtran}
\bibliography{evSDCP}

\vfill

\newpage
\appendix

\section*{Longitudinal Dynamics in the Spatial Domain}
To rewrite the vehicle dynamics in the spatial domain, we substitute the time-dependent variables with their spatial equivalents, where the independent variable becomes the travel distance \( s \) (or location index along the route, denoted as \( s_k \)). This transformation is achieved by utilizing the relationship between time \( t \) and distance \( s \), which is governed by the vehicle speed \( v \):  
\begin{equation}
\frac{ds}{dt} = v(t) \quad \text{or equivalently} \quad \frac{dt}{ds} = \frac{1}{v(s)} \,.
\end{equation}
The time derivative of the velocity, \( \dot{v}(t) = \frac{dv}{dt} \), can thus be expressed in terms of \( s \):  
\begin{equation}
\dot{v}(t) = \frac{dv}{dt} = \frac{dv}{ds} \cdot \frac{ds}{dt} = v(s) \frac{dv}{ds} = \frac{1}{2} \frac{dv^{2}(s)}{ds}.
\end{equation}

Combining all terms, the vehicle's longitudinal dynamics in the spatial domain become:  
\begin{equation}\label{eq:vehdyn-longitudinal-space}
\frac{m}{2} \frac{dv^{2}(s)}{ds} =  F_{\rm m}(s) - F_{\rm b}(s) - \frac{1}{2} \rho_\mathrm{a} C_\mathrm{d} A_\mathrm{f} v^2(s) 
- F_{\rm g}(s) \,.
\end{equation}  
Approximating the forces to be piecewise constants $F_{\rm z}(s) = F_{\rm z}(s_{k})$ for $s \in [s_{k}, s_{k+1})$, $\textrm{z}\in\{\textrm{m},\textrm{b},\textrm{g}\}$,
the \textit{exact} spatial-discretization of the dynamics~\eqref{eq:vehdyn-longitudinal-space} with zero-order hold inputs is obtained as
\begin{equation}
\begin{split}
\!
y_{k+1} & = e^{-\Gamma \Delta s_{k}} y_{k} \\
& \! + \frac{2}{m} \! \int_{s_{k}}^{s_{k+1}}\!\! e^{-\Gamma (s_{k+1}-s)}\!  \left( F_{\rm m}(s) \!-\! F_{\rm b}(s) \!-\! F_{\rm g}(s) \right) \! ds \\
& = A_{k} y_{k} \!+\! B_{k} F_{\rm m}(s_{k}) \!-\! B_{k} F_{\rm b}(s_{k}) \!-\! B_{k} F_{\rm g}(s_{k})
\end{split}
\end{equation}
where $\Gamma = \rho_\mathrm{a} C_\mathrm{d} A_\mathrm{f}/m$. The system parameters are given as $A_{k} = e^{-\Gamma \Delta s_{k}}$, and $B_{k} = 2(1- A_{k})/(m\Gamma)$. 
This spatial formulation facilitates the optimization of speed and energy consumption along the route and serves as the foundation for spatial domain planning strategies in electric vehicles.

\end{document}